**Toward an Effective Pedagogy of Climate Change: Lessons from a Physics Classroom**

*Vandana Singh, Department of Physics and Earth Sciences, Framingham State University*

**Introduction**

The crisis of climate change has been widely recognized by international organizations such as the Intergovernmental Panel on Climate Change and the World Economic Forum (IPCC, 2014; WEF Global Risks Report, 2017), as a significant threat to humankind and the biosphere, with some impacts already being felt and likely to worsen (IPCC 2014, 2018) in ways that we may not be able to entirely predict (Lenton et al., 2008b).  Additionally, climate change has been described (Levy & Patz, 2015; Weston, 2008) as a multifaceted justice issue, including intergenerational justice, since it disproportionately affects those who have contributed least to the creation of the problem, such as people of color, the economically disadvantaged, indigenous people, the rural poor, especially in the Global South, immigrants, including 'climate refugees,' and the young.  Young people, despite not having lived long enough to contribute significantly to climate change, are expected to bear a disproportionate share of the consequences, especially if they also belong to marginalized groups.  Unfortunately their education in high school and college is unlikely to have prepared them for the changes that are underway, in part because of systemic roadblocks in mainstream education (Kwauk, C., 2020).  If climate change is offered at all, it is either through specialized optional courses, or presented in various courses in a disjointed or piecemeal fashion, and, additionally, there is lack of adequate teacher training on the pedagogy of climate change (Plutzer et al., 2016).  My own experience indicates that fewer than 15% of students in a typical general physics college class at my institution have a basic understanding of climate science or climatic impacts on society.

Thus there is a compelling ethical argument for teaching students about climate change from middle school through their undergraduate years. In addition, making learning relevant to students' lives seems to be a factor in increasing motivation (Lazowski & Hulleman, 2016). Further, as recently discussed at a workshop at Columbia University's Earth Institute (Iyengar, R., 2020), the right kind of climate education may potentially serve as a climate mitigation tool.

However, despite the growing recognition of the need to teach climate change, the field of climate pedagogy is quite new, and beset with challenges. While the science of climate change is central to a comprehensive understanding of the subject, it is my contention that the impacts of climate change (ethical, societal, economic and biophysical) and the strategies for adaptation and mitigation call for a transdisciplinary approach informed by the best practices of transformational learning (Hoggan, C.D., 2018). Here I distinguish between multidisciplinarity (where multiple disciplines provide separate viewpoints on a particular subject), interdisciplinarity (in which two or more disciplines are combined in an integrative way) and transdisciplinarity (in which the distinction between disciplines is transcended to create a new way of thinking) (Leavy, 2011). I acknowledge the difficulty of developing interdisciplinarity in the current siloed education system, let alone transdisciplinarity; in fact many of the attempts toward a multifaceted approach to climate pedagogy stop at multidisciplinarity, where there is limited integration across disciplinary boundaries, especially between the natural sciences and the social sciences, and the natural sciences and humanities. Yet there is increasing recognition of the need to teach climate change and related environmental issues across the curriculum[1]

---

[1] See for example this excellent resource: https://tropicsu.org/

My motivation arises chiefly from three experiences: experimenting with pedagogy in college physics classrooms over the past eight years in a manner consistent with the ethos of transformational learning, developing an interdisciplinary case study on Arctic climate change (Author, 2015), and co-organizing a week-long workshop on an interdisciplinary approach to teaching climate change for middle and high school teachers in the US Northeast in 2017. My first intimation of the need for a transdisciplinary approach arose when I (a particle physicist by training) decided to learn and then teach the basic physics of climate change in a physics class in 2008. The students' reactions included despair, anger and apathy. Students also wanted to know what their lives might be like in a climate-changed world, and what they could do about the problem. This led to a realization on my part that teaching climate science in isolation was not enough. The naïve assumption that knowledge of climate change and its consequences is sufficient to prompt people to act – shown to be false (Chess & Johnson, 2007) - also did not hold in the classroom. In some cases, knowledge of the basic science of climate change caused the opposite of the intended effect – students felt helpless and disempowered, and fell into apathy. Thus the psychological dimension cannot be ignored in any effective climate pedagogy.

My approach to climate pedagogy arose in response to this and other experiences in the classroom, as well as the aforementioned case study, which included a trip to the Alaskan Arctic and interviews with scientists and Iñupiaq Natives in Utqiagvik, Fairbanks and Anchorage in 2014. The Arctic has been described as 'ground zero for climate change' because of the rapidity of sea ice melt and the anomalous rise in temperature. My Arctic experience made clear the importance of climate justice: the impact of climate change was greatest on the Iñupiaq people, who had not created the problem. In addition, the entanglement between indigenous culture, colonialism, economics, and oil drilling made obvious the complex nature of climate change impacts. The geophysical changes clearly affected culture and survival, and the responses of different stakeholders – oil companies, Inupiat, and others – also impacted the geophysical changes taking place. This led to the realization that any effective climate pedagogy must integrate the sciences with the human experience, and that climate justice must be an integral part of a holistic approach. In addition a basic introduction to complex systems science is central to understanding some of the apparent paradoxes and complications of climate science and climate impacts.

My third motivation for developing this approach comes from my experience co-conceptualizing and running an interdisciplinary week-long workshop for middle and high school science teachers the US Northeast in Summer 2017 in which some of the ideas of the interdisciplinary framework were first applied. Fifteen teachers participated in a range of inquiry-based experiences across disciplines, resulting in a wealth of data that continues to refine and improve my framework (Wade Institute for Science Education, 2017).

In this paper I describe this interdisciplinary approach that represents the first steps toward a transdisciplinary climate pedagogy, and which works around the limitations of disciplinary boundaries as they manifest in most high schools and colleges in the U.S. Since it is not always possible for most students to take a course that covers multiple aspects of climate change, the challenge is to introduce it in various required courses in a manner that is relevant to the discipline, without losing the interdisciplinary aspect. This paper describes the development of such an approach in the context of general physics college courses ranging from 'physics-for-poets' type courses to calculus-based physics, with particular emphasis on the former. I describe the challenges that emerged from this experience and introduce the first steps toward a transdisciplinary framework for addressing these challenges. I suggest that such a framework can be adapted for any general science course in high school and college (with the possibility of extension to non-science courses) to introduce climate change in a subject-relevant manner without losing its multifaceted aspects. It is to be noted, however, that this framework is only one among many possible, and I hope that educators around the world are inspired to come up with their own transdisciplinary frameworks that arise from their specific contexts. Therefore I consciously eschew any reductionist 'one size fits all' claims about this framework. Since the pedagogy of climate change is still in its infancy, further discussion, exploration, refinement, modification, and testing of these ideas as well as the generation of new ones, is crucial.

**What is an effective inter-to-trans-disciplinary pedagogy of climate change?**

Here I clarify what I mean by an effective climate pedagogy. What do we want students to take away from their study of climate change? This is not a simple question. One possible answer from the perspective of a science educator is that we need them to understand the science, and that is where our responsibility ends. We might assume that knowledge will empower students to act, but studies show that knowing how serious the implications are does not result in action, rather it can impede action (Chess & Johnson, 2007). In my experience, teaching climate change science without inter-/trans-disciplinarity embedded in a transformational learning framework, results in many students experiencing anger, despair, denial, and apathy, often resulting in a lack of motivation to act. The pervasive argument that scientists should simply do the science (and by implication, teach just the science) is no longer tenable in the era of climate change and related crises.

I propose that an effective pedagogy of climate change in a physics (or other science) classroom is one that:

a) equips the student with a fundamental understanding of the basic science, impacts, and evidence of climate change, including its complex, nonlinear nature, as well as the future projections based on various scenarios – *the scientific-technological dimension*

b) enables the student to understand societal and ethical implications of climate change (climate justice), leading to intersections with economic, cultural, human rights and sociological issues; to understand how climate change is related to other major social-ecological problems and to critically examine proposed climate solutions from a climate justice perspective – *the transdisciplinary justice dimension*

c) enables the student to undergo an epistemic shift (Mezirow, J. & Taylor, E.W., 2009), and thereby see the climate crisis as a symptom of a social-scientific framework or paradigm; so as to understand and articulate the need for new social-scientific frameworks in order to usefully engage with the crisis - *the epistemological dimension*

d) inspires students to explore their own response to the crisis, as well as their agency, and encourages them engage with social-environmental problems in society – *the psychosocial-action dimension.*

Although the above are at the classroom level rather than program level, they further the development of four of five key competencies identified for sustainability education: systems-thinking, anticipatory, normative and interpersonal competencies (Wiek, A. et al., 2011). Additionally, this pedagogical framework seeks to address the challenges listed in the next section.

**Challenges to Teaching Climate Change**

At the macro level, scholars have identified five major roadblocks to 'quality education in the time of climate change (Kwauk, C., 2020): low priority for ecoliteracy, lack of a radical vision for education, a problem of definition and scope in education for sustainable development (including a narrow focus on the science, and an approach counter to the principles of transformational learning), monitoring and accountability mechanisms geared to passive progress, and finally, lack of systemic support for teachers to become 'change agents for sustainability.' These also manifest in the microcosm of the classroom, although they do not map onto each other one-to-one. Based on my experience and on the literature, I have identified the following barriers.

a) Knowledge pollution and ignorance: There is significant conceptual confusion and ignorance among the U.S. public and the world at large about anthropogenic climate change (Lee et al., 2015; Leiserowitz, A. et al., 2010), leading to misconceptions such as the confusion between the climate crisis and the ozone problem ("the hole in the ozone layer is heating up the planet.") (Leiserowitz, A. et al., 2010). Unreliable sources in the media and popular culture magnify misconceptions. For example studies (Cook et al., 2016) show that there is overwhelming consensus among climate scientists that climate change is happening and is primarily due to human activity. But among the American public just over half believed this to be the case in 2017 (Ballew et al., 2019). Political partisanship on the issue and the reported funding of skeptic groups by fossil fuel companies (Cook, J. et al., 2019) has further obscured the issue in the public mind. Some confusion also arises because of the public's lack of understanding of how science works. Science by its very nature is provisional and subject to constant revision and correction. Climate science has specific challenges that result in some uncertainties that are inherent and others that represent a lack of knowledge (Risbey & O'Kane, 2011). However the basic science of climate change, resting on thermodynamics, the physics of greenhouse gases, and the carbon cycle, is indeed reliable "with high confidence" (IPCC, 2014). But, because of unfamiliarity with how science works and b) above, reports of uncertainties in climate predictions can lead the public to conclude that the reality of anthropogenic climate change is in question.

b) The challenge of inter-/transdisciplinarity: Climate change is a phenomenon at the intersection of physics, chemistry, biology, economics, sociology, psychology, and indigenous rights, to name but a few areas – the ultimate 'wicked problem.' Our siloed system of education does not easily allow space for a truly inter/transdisciplinary exploration of climate change. Disciplines have developed specific lexicons, paradigms, and frameworks that may not be easily translatable across boundaries; hence transdisciplinary scholarship is a relatively new field (Brown, V. A. et al., 2010; Leavy, 2011).

c) Psychological barriers – the psychology of climate change is still in its infancy but recently (Running, 2007) it has been suggested that learning about climate change results in emotional trauma that includes denial, anger and despair. This is consistent with my teaching experience. Climate change and its implications are frightening. The dilemma is that while we, as educators, are obligated to tell the truth, the truth can be an overwhelming emotional burden. Science educators are not equipped to handle the affective impact of their teaching upon students. Being aware that climate change is an emotionally fraught issue (APA Taskforce on Interface Between Psychology and Global

Climate Change, 2009), we can apply ways to work through this aspect that allows the student to engage meaningfully with the subject.

d) Onto-Epistemological Barriers – modern industrial civilization encourages short-term, linear, individual-centric, localized thinking, aspects of the dominant paradigm that does not fit the complex, entangled, spatially and temporally long-range nature of the climate crisis. Thus 'solutions' arising from the same paradigm or worldview that brought us the climate crisis are unlikely to be truly effective. Multiple scholars of transformative education have pointed out that an epistemic shift must undergird any effective pedagogy of social-environmental problems (Boström et al., 2018; Lotz-Sisitka et al., 2015; Macintyre et al., 2018; Mezirow, J. & Taylor, E.W., 2009; Odell et al., 2020; Sterling, 2011).

e) Faculty Training and Institutional Barriers

The above challenges make it imperative for faculty in all disciplines to receive training in transdisciplinary and transformative education methodologies. This would entail collaboration between far-apart disciplines such as physics and sociology, or biology and literature. The rigid structural barriers in educational institutions effectively wall off disciplines from each other; there is generally little recognition or appreciation for the value of cross-disciplinary training and collaboration. Thus educators wishing to teach climate change across disciplines have to teach themselves. In my own case, my training is in particle physics, so I had to take the initiative to learn climate science (an ongoing journey), and to ask myself this crucially important question: from the vast and rapidly developing field of climate science, what concepts can be considered essential to a lay but sophisticated understanding of the crisis? Since I could only teach climate change in the context of a physics class, I had to figure out these essentials and learn how to connect them with existing course topics. I also had to learn from sociologists, economists, anthropologists, and scholars of environmental humanities, among others, and integrate key ideas from those disciplines into my framework. From the logistical challenge of teaching climate change in a classroom devoted to another subject (even with a transdisciplinary approach) there emerged a significant problem: that of *piecemeal learning*, where climate topics appear at irregular intervals during the course in disconnected chunks. Such piecemeal learning would prevent an integrated, comprehensive 'big picture' understanding of the material, and undermine the intended purpose.

**Designing an Inter-/Trans-Disciplinary Framework of Climate Pedagogy**

Below I describe an interdisciplinary approach to integrating climate science and justice in a course not specifically devoted to the climate crisis that is inspired and influenced by a transdisciplinary ethos. (A truly transdisciplinary approach would have a different starting point – a case study or a narrative, for example, that from the start transcends disciplinary barriers. This is ongoing work).

1. Planning the Course:
a) The instructor explores where climate change intersects with the subject matter of the class. For a general physics classroom, there are several intersections with climate physics. We connect those climate change topics with the subject matter and work out a schedule.

b) Based on the four dimensions of the effective climate pedagogy described above, the instructor considers what is missing from the plan so far, and works on incorporating the remaining dimensions of effective climate pedagogy.

c) The instructor then comes up with a plan that *integrates* the four dimensions of effective climate pedagogy, thus avoiding 'piecemeal learning.'

2. Climate change is introduced before the semester begins and discussed on the first day of class, as described in more detail below. Students are guided in the first week toward constructing a holistic visual tool, Figure 1, which, in each iteration, acquires more layers of meaning and complexity as students gradually deepen and complicate their understanding of key ideas as they revisit them or see them in the light of new knowledge. It therefore serves as a scaffolding device, referred to whenever a climate change topic is discussed, so that these topics are seen as part of a holistic framework. The key features include three cross-cutting, inter-related, overarching concepts that I call *meta-concepts*.

3. Making it Relevant: Woven throughout this framework is an attempt to make climate change feel personal and relevant to students' lives. Therefore room is made through the course for exploring the affective impact of climate change, bringing in stories from students' lives, and discussing climate justice. A class project that is transdisciplinary in nature and has a service component further reinforces these ideas. In addition, discussion of solutions is not limited to science and technology, or to individual actions, but includes interventions by indigenous peoples, climate movements, and alternate paradigms. A classroom culture that is collaborative and based on trust is essential for the success of this approach, as elucidated below.

**A note on Teaching Approach and Method**

My teaching approach is inspired by the work of Carol Dweck on mindsets (Dweck, C.S., 2006), and of Ken Bain on the Natural Critical Learning Environment (Bain, 2004; Bain, K. & Zimmerman, J., 2009) as well as Embodied Learning (Euler et al., 2019). More recently I have discovered that my approach is consistent with the ethos of transformational learning (for an overview, see (Hoggan, C.D., 2018)). I have been practicing this approach and improving on it for the past eight years, in all levels of undergraduate general physics courses, in particular Physics, Nature and Society, a laboratory physics course for non-science majors. Some of these techniques have also been employed in a First Year Seminar on Arctic climate change. I start with posing an overarching question or theme for the course before the course begins, which is sent out as an invitation to students ((Bain, 2004). For Physics, Nature and Society, the theme has consistently been climate change. The courses are taught with an active learning focus - interactive lectures are interspersed with think-pair-share exercises, group work, student exploration with laboratory equipment, embodied learning in the form of 'physics theater,' whereby students work in groups to enact physical principles, chances for students to hypothesize without worrying about being wrong, second chances on certain exams, and a focus on the individual student – i.e. helping *every* student toward excellence through extra help and encouragement. A consistent effort is made to build trust between instructor and students, and to create an environment where students feel psychologically safe and valued, so that they are equipped for the intellectual challenge. The details of this approach are reported elsewhere (Author, Manuscript in Preparation) and our pilot study indicates improvement in confidence, interest and performance among a majority of students in our classes.

Since teaching climate change is beset with challenges additional to those that we encounter in the general physics classroom, it is important to employ these best practices for teaching in a physics classroom; I do not believe the traditional educator-centered lecture format is conducive to an effective pedagogy of climate change.

This leads to an important observation and drawback of my approach. Unlike a clearly defined physical system in a laboratory, with a limited number of independent variables, the system consisting of students and educator is by its very nature complex (Brown, V. A. et al., 2010). To show the efficacy of any pedagogical approach, we need to control multiple variables – however, in the classroom, these variables are not independent. For example the interaction between one pair of students is not the same as the interaction between another pair of students (unlike the case of molecules in an ideal gas) and this also applies to teacher-student interactions, even if the teacher tries their best to treat each student on an equal footing. How the social-psychological aspect of learning affects the intellectual and cognitive aspect is a subject of considerable research (Yeager et al., 2013) and we know now that one affects the other. Pedagogical research often involves showing correlations between student success and some intervention, and then attempting to establish causation through a theory or theories of learning. However there are multiple reasons why I do not take this approach. One is that in order to effectively teach a subject as inherently transdisciplinary as climate change, we need to approach it through a framework or a philosophy, rather than a simple intervention or two, and therefore we are dealing with many more interdependent variables than is usual for a pedagogical experiment in a normal physics classroom. For logistical and ethical reasons I do not set up a control group. Additionally my classrooms tend to be small (N varies from 9 students to 26 students) so I cannot draw any grand conclusions from my results. Although my work developing this approach was carried out over a period of eight years, it does not make sense to combine the class N values into one set, because different features of this interdisciplinary framework were developed at different times during the seven-year period. Nor is it possible to treat my study like a collection of pilot studies because my approach is iterative: the success or failure of one aspect of this framework is likely to depend on the success or failure of another that might have been developed at a different time – in other words, the situation is neither linear or static.

An additional reason is that my pedagogical approach is one of 'participant-observer,' that is, I am (inevitably) a part of the system I am studying. A deep engagement with every student, including many hours of interaction outside official meeting times, allows me to sense, observe, and respond to cognitive and emotional changes in the student, which is also a source of useful qualitative information.

Finally, while the intent of this approach is to provide the conditions for students to experience an epistemic shift (a central concept in transformational learning) (Boström et al., 2018; Macintyre et al., 2018; Mezirow, J. & Taylor, E.W., 2009; Sterling, 2011), this is not possible to measure through conventional means. Such an epistemic shift might take place in a series of stages, and its effects may not be apparent until long after the class is over. The lack of transformative pedagogy in other classrooms could erode the impact of this pedagogy over time.

Students' responses (affective and cognitive) to learning about climate change were collected via discussion questions, tests, homeworks and exams, as well as anonymous surveys. Student responses and changes therein were also ascertained anecdotally in nonmeasurable ways through constant and deep engagement with them. I refer to all these responses in the course of this paper, but I do not perform detailed statistical analyses for the reasons elaborated above. Thus this paper is primarily qualitative, descriptive, and tentative in its conclusions, its major purpose being to make a case for inter-/transdisciplinarity in teaching climate change in a science classroom through the methods of transformational learning, by presenting the development of one framework (among many possible), and thereby to invite responses, critiques, ideas and further experimentation.

**Intersections between Climate Science and Topics in a General Physics Course**

This section may be omitted by those not teaching physics courses, although a glance at the table of topics and intersections with climate science may be helpful, especially the entry on 'Climate Week.'

The basic physics of climate change intersects with fundamental physics course topics in five areas – fluids, the electromagnetic spectrum, the idea of thermodynamic equilibrium and energy balance, the concept of resonance in oscillations, and the notion that electromagnetic waves are generated by oscillating electric dipoles. Depending on the course, these areas and their relation to climate physics can be discussed in varying levels of detail. Instead of waiting until all these topics have been covered, climate physics can be introduced in stages (preceded by a short contextualizing discussion; see next section): in relation to a) oscillations and resonance, which explain in part why carbon dioxide is a greenhouse gas, then picked up b) immediately after the EM spectrum is studied, to introduce the greenhouse effect, and c) discussed explicitly thereafter as described below, putting the pieces together and introducing new ideas in an integrated manner. If generation of EM waves via oscillating dipoles is discussed later in the course, students can then better understand why carbon dioxide is a greenhouse gas. This approach, limited as it is to general physics courses, leaves out (except for a crucial short exploration of complex systems) interesting and important details from the Earth Sciences such as atmospheric structure, the role of winds and ocean currents, the absorption spectra of greenhouse gases (which are studied briefly but not elaborated on), and the detailed interaction of various earth systems to produce climate, which are best explored in higher level specialized courses. However the 'big picture' is introduced right from the start, via Fig. 1, which is revisited each time a climate change topic is addressed.

The table below shows how standard physics concepts intersect with climate physics, in the order that these concepts are introduced in this study. Note that the choice and depth of subtopics depends on the level of the course being offered, but the Greenhouse Effect, the Carbon Cycle, and sea level rise as well as impacts and projections are discussed in every course at the appropriate level of detail.

**Implementing the Interdisciplinary Framework:**

1. A Contextualizing Discussion –

Before classes start, the topic of climate change is introduced via email as an invitation to the student to participate in an exploration of the climate crisis, with the 'promise' that the physics course will give them the tools to make sense of this urgent real-world problem.

Within the first two weeks of class, students are taken to the University Planetarium for an immersive experience in which satellite images of the Earth are presented. I provide them with blank copies of Fig. 1 (without the text). With some encouragement, students identify the Earth's five subsystems as conceptualized by Earth scientists: atmosphere, hydrosphere, cryosphere, geosphere and biosphere. We then look at the image of the planet at night, where the human impact on the planet becomes obvious, leading us to the sixth proposed subsystem: the Anthroposphere. I then ask the students to hypothesize why the Earth's land areas are not equally illuminated. An overlay of information about human population tests student conjectures about population density, and makes inequality evident as a key complication. We are thus able to interrogate the appropriateness of the term 'Anthropocene' by noting that not all humans have the same impact on Earth systems, and these humans tend to live in poorer and formerly colonized nations. We finally settle on Modern Industrial Civilization as the more accurate (if imperfect) alternative to 'Anthroposphere.'

In early classroom discussions of climate change, students are invited to raise questions whose answers will eventually lead to a comprehensive understanding of basic climate change. Typical questions that arise include: "How do we know whether climate change is real or not?" "Can't it be due to natural causes?" In the last two years, for reasons that will become apparent, students have also been asked to share their feelings, if any, about climate change. Responses range from "I don't care," to "I'm freaking out about it." These are also noted and students are promised that all these questions and responses are valuable, and that the answers will emerge from their study of physics within an interdisciplinary framework.

2. The Meta-Concepts

After the first climate science topic (blackbody radiation and planetary equilibrium temperature) in the semester schedule has been introduced, students are introduced to the Meta-concepts via Fig. 1, but without much detail as yet. As the course progresses, Fig. 1 is used for each section as a unifying tool – for example, after the students study the Greenhouse effect or the carbon cycle, they can use Fig. 1 to identify the subsystems interacting among each other and with the sun. An embodied approach is employed, in which students work in groups to enact physical phenomena, including the modes of oscillation of the carbon dioxide molecule. This is based on a body of studies on the multiple efficacies of embodied learning (Euler et al., 2019; Solomon, F.C. et al., 2019).

The Meta-Concepts are introduced via the following sentence that accompanies Fig. 1:

*Global Climate is a **complex system** based on the non-linear interactions of six major subsystems and the sun, that can result in **balance/imbalance** depending on whether (or not) the earth systems are operating within certain **planetary limits or boundaries**.*

I elucidate these in detail below. The motivation for the meta-concepts is one, to enable a holistic, rather than piecemeal understanding of climate change, two, to create a conceptual structure that, within the classroom culture created through transformational learning, potentially allow us to surmount the remaining four of the five barriers mentioned earlier, thus aspiring to the four dimensions of an effective pedagogy of climate change. The scientific basis of each meta-concept is intended to create a misconception-proof understanding of climate essentials, thereby addressing the barrier of Knowledge Pollution. Specifically the following questions are addressed:

a) What is the difference between climate and weather? What is climate change and what is the evidence that it is happening?
b) What does the Earth's paleoclimate history tell us about current climate change and the 'human footprint?'
c) What are the causes of climate change, and how does it connect to other social-environmental issues?

d) In what ways is climate a complex system and what are the implications (in terms of impact and actions)?

e) What are the differentiated human causes and impacts of climate change?

In a small study scholars in the UK (Hall, Brendan M., 2011) have identified these threshold concepts that arise in the interdisciplinary teaching of climate change: uncertainty, the geological perspective, the paleoclimate toolbox, climate science fundamentals, and modelling and scenarios. "A threshold concept can be considered as akin to a portal, opening up a new and previously inaccessible way of thinking about something." (Meyer, J. & Land, R., 2003). These are closely related to the idea of 'troublesome knowledge' (Perkins, 1999 as quoted in the above) which is knowledge that is 'troublesome' in some way for the learner, requiring, for example, unfamiliar ways of thinking, tacit or inert knowledge, or unusual structural complexity. Threshold concepts tend to be 'troublesome' in multiple ways; in the climate context seeing the Earth as a complex system is especially challenging. The meta-concepts of balance, limits and complexity embrace all of these troublesome threshold concepts, with special emphasis on: uncertainty, the geological perspective, and climate science fundamentals. In the context of a classroom culture engendered by the teaching philosophy described earlier, these threshold concepts may potentially be portals toward what transformational learning theorists call an epistemic shift (Mezirow, J. & Taylor, E.W., 2009).

The barrier of transdisciplinarity is addressed by centering justice within the meta-conceptual framework. Justice issues are raised while studying each meta-concept, and allow us forays into such varied disciplines as economics, sociology, indigenous rights, and the role of the humanities. Our study of complexity and alternative epistemologies, arising after our discussion of paradigm shifts in science early in the course, allows students to see that there are multiple ways of living and being in the world, and that the way they take for granted is not the only way. This addresses the barrier of onto-epistemology. Psychological barriers are addressed in three ways: one, through the embedding of the meta-conceptual framework in a transformational learning approach that gives customized support to each student, two, by giving space for students to express how they feel about the climate crisis and supporting them emotionally, and three, through engagement with real-life climate justice stories and a classroom project, described later on.

1. Balance/ Imbalance

   Here 'balance/ imbalance' refers to a steady state/ non-steady state, which terms can be used instead for a higher level class. Conditions of balance and imbalance can arise due to natural factors alone, as Earth's paleoclimate history demonstrates, but current climate change (a departure from relatively stable conditions of the Holocene, and also a departure from the oscillations between the glacial and interglacial periods of the last million years ) is primarily due to human activity (IPCC, 2014).

   This meta-concept contains the heart of the basic science of climate change and allows for many opportunities for critical thinking and rich conceptual discussions. Outlined below is the sequence in which we introduce topics through the semester as indicated in Table 1. We use iconic graphs and visuals from reliable sources on the web to anchor the key ideas. In a non-science-majors physics class these are sufficient; in higher level classes these can provide the conceptual foundation for a more quantitative exploration.

   a) We study the sun's radiation spectrum (blackbody curve) and compare it to that of the Earth. Students note that the sun emits mostly visible and high frequency infra-red waves whereas the Earth's radiation is in the low frequency IR regime. Depending on the level of the class, Planck's formula and Wien's law may be introduced mathematically as well as conceptually.

b) The greenhouse effect is introduced. In a non-science-majors classroom this can be done purely conceptually; in higher level classes we additionally use the Stefan-Boltzmann law applied to the Earth in thermal equilibrium to calculate the surface temperature of the Earth (and Mars and Venus for contrast) with and without greenhouse gases in the atmosphere (as discussed, for instance, in the very useful introductory book by Archer (Archer, D., 2010)). The difference in temperature indicates the greenhouse effect. Students look at Figure 1 and discuss which of Earth's subsystems participate in the natural greenhouse effect. This use of Fig. 1 as a unifying scaffolding for climate change topics is continued throughout the semester.

c) We discuss the greenhouse effect and the Earth's energy budget, introducing some numbers (such as the total energy flux from the sun, how much is reflected by clouds and the ice caps, how much is absorbed and re-emitted as low frequency IR).

d) Carbon dioxide is a greenhouse gas because of the way the molecule responds to infrared radiation. Although the role of carbon dioxide as a warming gas is mentioned during the discussion of the greenhouse effect, the details of the molecule's vibrations are introduced after oscillations and resonance are covered later in the semester, as per the sequence in Table 1. Students visit a chemistry lab at our university where a physical chemist leads a demonstration with an IR spectrometer, revealing the absorption spectrum of carbon dioxide. A model of the carbon dioxide molecule built by colleagues in our Chemistry and Food Science department vividly demonstrates the three main vibrational modes of $CO_2$ and the significance of the bending mode. These demonstrations (Bell & Marcum, 2018) help make the classroom discussion of carbon dioxide as a greenhouse gas clearer and more tangible. When electromagnetism is studied later in the semester, more details about the bending mode of $CO_2$ may be added, but this is not necessary for a non-science-majors class.

e) Students study the graph of carbon dioxide atmospheric levels in ppm over the last 800,000 years (Figure 2) - for example as available on this website (US Global Change Research Program (Archived), 2009) - and discuss it in small groups and share their observations. The students note that $CO_2$ concentration, while 'oscillating' between 200 ppm and 290 ppm, does not exceed the latter number until more recent times (the concentration since the industrial revolution can be shown in a more detailed graph), after which there is a sharp spike. We then introduce the three 'dials' that control the Earth's temperature: albedo, changes in the sun's activity, and atmospheric greenhouse gas concentrations. Milankovich cycles (where the Earth subsystems – primarily cryosphere, hydrosphere, and atmosphere, interact with the sun in the context of regular perturbations in the Earth's orbit, generating the glacial and interglacial periods) are briefly mentioned to explain the rise and fall of $CO_2$ levels in this graph.

f) In the context of this graph we can introduce the idea of balance (or steady state) and imbalance with reference to $CO_2$ levels, which will later be examined more carefully when we discuss the carbon cycle. It is worth emphasizing that imbalance (when $CO_2$ levels are changing instead of staying steady) can also result from natural factors (as evidenced by the Earth's paleoclimate record), thus countering the popular misconception that all natural states without human interference are 'in balance.' 'Imbalance' can describe oscillatory behavior (glacial-interglacial cycles) in which $CO_2$ levels and temperature go back and forth about an average, over the past million years. It is worth noting that balance/imbalance also depend on timescale – for instance an expansion of the time axis in the above graph will reveal periods during the glacial/ interglacial cycles where carbon dioxide levels did not change very much. The complex relationship between the internal dynamics of the Earth system and external perturbations causes temperature and $CO_2$ to rock back and forth in sync about a mean (which results in the drastic difference between a glacial period and an interglacial period, a useful illustration of the fact that a small change in $CO_2$ has a disproportionate effect on Earth's climate). But with the onset of the industrial revolution we see a different kind of imbalance – a steep rise in $CO_2$ levels – in the past 150 years, with no possibility of return within the timescale of the glacial-interglacial cycle (Archer et al., 2009). It is useful also to consider the rate at which atmospheric $CO_2$ levels are rising (over 2 ppm/year in the last few years); in the past 60 years the rate of increase is 100 times greater than the rate of change at the end of the last ice age, 11000 to 17000 years ago (Lindsey, R., 2020). The sharp rise in both $CO_2$ and temperature since the industrial revolution, points to the burning of fossil fuels as a *likely* factor, but without additional information (such as the known warming effects of $CO_2$) it establishes a correlation and not a cause. This is a good place to discuss how science actually works – evidence is built from multiple sources that converge, and cause is established via well-tested experiments, observations and models that relate the variables in a manner consistent with physical laws. Time permitting, different kinds of uncertainty and confidence levels as applied to climate change ( the epistemology of the field is still developing) may also be discussed in higher level classes (IPCC, 2014; Lewis, S. & Gallant, A., 2013; Risbey & O'Kane, 2011).

g) We next introduce natural carbon cycle via a series of images[2] that are easily available on the web. These indicate carbon sources and sinks, including rates and timescales. Students examine the diagrams in small groups and share what they've learned and what questions arise. With the help of a carbon cycle activity described in Box 1, they note that natural carbon sinks and sources working at roughly the same rate results in a constant level of $CO_2$ in the atmosphere, and that the extraction and burning of fossil fuels – which, being buried underground, do not participate in the carbon cycle except on very long timescales – introduces a new source of $CO_2$ into the atmosphere. Despite being small compared to some of the natural flows of carbon, this new carbon source, along with the destruction of sinks (such as tropical rainforests, the second largest 'source' of anthropogenic greenhouse gas emissions (Pendrill et al., 2019)) throws the carbon cycle out of balance, allowing $CO_2$ atmospheric concentration to rise. Here I find that the bathtub analogy (see for instance the diagram here (EPA (Archived), 2014) and the abovementioned carbon cycle activity help make concrete the difference between a steady state and a non-steady state.

h) We discuss the evidence that increasing $CO_2$ in the atmosphere is indeed due to 'human activity'. The key points are:
   - Correlation between sudden rise of $CO_2$ levels and the industrial revolution post-1750
   - Known ability of $CO_2$ to 'trap' terrestrial infra-red radiation (Greenhouse effect)
   - Evidence from carbon isotope ratios (relevant for more sophisticated classrooms)
   - Evidence from climate models – models run with only natural factors show that we should be on a slightly cooling trend, which, of course, does not match observations.

---

[2] See, for example, images here: https://www.sciencedirect.com/topics/earth-and-planetary-sciences/global-carbon-cycle and https://science.nasa.gov/earth-science/oceanography/ocean-earth-system/ocean-carbon-cycle

When the human contribution is added to the climate models, they track observed temperature rise quite well (EPA (Archived), 2014).

> **Box 1: A classroom activity to teach Carbon Cycle Disruption in the Climate Context**
>
> How do we convey the difference between a steady state (Balance) and a non-steady state (Imbalance) to students learning about the anthropogenic disruption of the carbon cycle in the context of climate change? Rates of change are difficult to teach; yet this is a fundamental concept without which a student's grasp of key climate science basics will be incomplete and may lead to misconceptions. We describe below a classroom activity that helps clarify the changes in atmospheric carbon dioxide composition over 800,000 years through the present day. The idea is for students to understand the significance of the graph of these changes (Figure 2) in the context of the anthropogenic influence on the carbon cycle. The inspiration for this activity is the bathtub analogy for the atmospheric carbon dioxide concentration referenced in the text.
>
> First, we draw, with a marker or masking tape, a rectangle on the classroom floor, large enough to accommodate 5 students (the size would depend on the classroom size; we recommend that the rectangle should be large enough that a third of the class can stand in it at a comfortable distance apart at any time during the activity). At one end of the rectangle we have a gap in the masking tape that we label the entry point, and a similar gap at the other end labeled the exit. Students line up in front of the entry point. I then ask for two volunteers to be the ushers, one to usher students into the entry point of the rectangle, the other to usher students out of it. Each usher waits at the ready outside the entry and exit points respectively. Another student or two are designated as observers who will note, at the appropriate time interval, how many students are in the rectangle.
>
> The roles of the students are explained briefly. Except for the ushers and observers, the remaining students are going to play the role of carbon dioxide molecules in the atmosphere. The rectangle symbolizes the atmosphere, and the ushers are respectively the carbon sources and sinks, already somewhat familiar to the students from their first look at a diagram of the carbon cycle.

> I then tell the students that the ushers will attempt to work at the same rate to bring students in and usher them out of the rectangle. The activity begins. After a little while it becomes obvious that when the first usher is putting students into the rectangle as fast as the second usher is bringing them out, there is generally a constant number of students in the rectangle. The student observers can, for example, note every 30 seconds (or similar convenient time interval) there are always 5 to 6 students inside the rectangle. If needed, these students can plot the number of students in the rectangle versus time on a giant sheet of graph paper or on the whiteboard.
>
> Now the first usher (at the entry point) representing the Earth's carbon sources declares that they are going to increase the speed at which students are ushered into the rectangle, while the second usher at the exit point maintains the original speed. As the activity proceeds, we see that every time interval, the number of students inside the rectangle goes up. The student observers will then note that their flatline graph is rising. This is a non-steady state (imbalance).
>
> After the activity students are first encouraged to articulate their observations. Then we go back to Fig. 1 and Fig. 2, and apply what has been learned in the activity to the problem of the carbon cycle. We use think-pair-share or group discussions before opening it up to the class as a whole.
>
> A misconception that might arise is that if the rectangle is not large enough, students might incorrectly conclude that the atmosphere is" getting crowded with carbon dioxide." In fact $CO_2$ concentration in the atmosphere is very small, around 0.04 %. If, during the study of the climate crisis, students are taught about carbon dioxide's atmospheric concentration relative to other gases, and if we emphasize that during this activity we are ignoring the far more abundant gases such as nitrogen and oxygen, this misapprehension can be avoided.

i) When $CO_2$ concentration rises in the atmosphere, it is worth discussing how long it is expected to stay there. One of the diagrams examined by the students shows the carbon cycle operating over multiple timescales, the longest of which is the carbonate-silicate cycle with a time period of hundreds of thousands to a million years. Models indicate (Archer et al., 2009) that while about half the $CO_2$ humans release today will be taken up by the land biosphere and the surface oceans in a matter of a century or two, a significant fraction (20 – 35%) will persist in the atmosphere on a scale of centuries to millennia. The longevity of carbon dioxide in the atmosphere allows us to discuss the idea of intergenerational justice and the responsibility that those of us alive on Earth today have to several future generations of humans and nonhumans who will have to deal with the consequences of our actions.

In the context of the metaconcept of 'Balance/Imbalance,' we also discuss how the imbalance in the carbon cycle has arisen. The connection between the industrial revolution, colonialism, and environmental degradation (Murphy, J., 2009) and the continuing entanglement of GDP and fossil fuel use can be brought out more explicitly here. Along with the discussion on responsibility to future generations, this allows us to reconnect with our preliminary Contextualizing Discussion, where an examination of Earth images made social inequality on a global scale evident. Thus we disambiguate 'human' here and throughout the semester by asking if all humans are equally responsible for climate change. This allows students to critically examine the widely used term 'anthropocene' and introduces more deeply the idea of climate justice. Climate justice is explored more fully during Climate Week in the schedule above, but the concept is first elaborated when we discuss the 'human' attribution of climate change, and is woven through all subsequent discussions.

2. <u>Planetary Boundaries and Limits:</u>

Having established the meta-concept of balance/imbalance – with reference to both $CO_2$ sources and sinks, and thermodynamics - it becomes clear that any physical system that exists in a steady state (in terms of $CO_2$ concentration and temperature) must therefore have limits – boundaries beyond which the system departs from a steady state (the transition from balance to imbalance). The system's response may be slow and gradual, or abrupt; it may be reversible or irreversible on human timescales. Not all steady states may be conducive to modern industrial civilization, or even to the existence of humans, so our discussion of planetary boundaries is biased toward balance conditions that are optimized for human wellbeing (but not necessarily modern industrial civilization as it is today). (It is to be noted that there is no one idea of 'human wellbeing,' therefore it is important to invite pluralistic conceptualizations of the term (Kothari, A. et al., 2019).) Planetary boundaries and the notion of a safe operating space for humans (Steffen et al., 2015) is an ongoing area of research.

The proposed nine boundaries include climate change, biodiversity, and ocean acidification. Crossing these boundaries can imperil the conditions that make life (as we know it) possible; the authors of the research at Stockholm University conclude that we have already crossed four of these nine boundaries. Note that a recent paper (Steffen et al., 2018) considers the Earth system to be at a critical point, a fork in the road past the glacial-interglacial cycles, where future action or inaction will determine whether we end up with a human-nurtured stabilized Earth, or one that careens uncontrollably toward a hothouse state not seen for millions of years of Earth history.

Another way in which the Planetary Boundaries concept is useful is to demonstrate that climate change is not the sole ecological problem facing the biosphere. Any inter-to-transdisciplinary approach to teaching climate change should contextualize it within larger issues. Discussion of planetary boundaries is an opportunity to briefly introduce other violations of these boundaries – the imbalance in the Nitrogen cycle, for example, as well as the dire situation of biodiversity loss: a million species threatened with extinction, as reported in a recent UN publication (Intergovernmental Science-Policy Platform on Biodiversity and Ecosystem Services, 2019) and the consequences of land system change. The fact that these boundaries are not independent allows for rich discussion as to how, for example, biodiversity loss might exacerbate climate change, and vice versa. This allows students to appreciate that climate change is part of a much larger complex of problems, and therefore technological 'solutions' for climate change alone that ignore the larger, deeper issues are inadequate at best. Such a horizon-widening discussion is essential because it contextualizes the climate crisis within a certain paradigm through which these other crises also arise, and points the way toward examining their common root causes and features.

The idea of limits as applied specifically to climate change is introduced via this question: what is the 'safe upper limit' of $CO_2$ concentration in the atmosphere? Or "what is the 'safe upper limit' for the Earth's global average surface temperature?" In 2009 governments and scientists agreed at the UN Copenhagen meeting that 2℃ was the ceiling beyond which global temperature could not be allowed to rise. This was controversial, as there is no cut-and-dried formula for 'safe upper limit.' In the UNFCCC meeting in Paris in 2015 the aspirational goal was set to 1.5℃, (Gao et al., 2017) although the world is likely to be heading toward a greater than 4˚C rise by 2100 if no action is taken (Tollefson, J., 2020). However some important research in 2011 (Leaton, J., 2011; Meinshausen et al., 2009) estimated that only 565 Gigatons of $CO_2$ (in 2011) could be permitted into the atmosphere before reaching a 2˚C rise, whereas we already have enough global fossil fuel reserves that, if burned, would release about five times this 'safe' amount. Currently fossil fuel companies are spending billions of dollars looking for new and increasingly hard-to-extract deposits of fossil fuels, despite the fact that current reserves are already in excess of climate limits (Carbon Tracker Initiative, 2017; Grant, A. & Coffin, M., 2019). Seen within this metaconceptual framework, such 'business-as-usual' activities stand out as senseless and dangerous.

It is helpful to introduce the key findings of the 2018 IPCC Special report comparing a 1.5 °C rise with a 2 °C rise (IPCC, 2018). Given that our current 1 °C rise already presents us with substantial threats to human and natural systems, a 1.5 °C world, while a lot more challenging, is far preferable to a 2 °C world, which would likely be catastrophic.

A critique of the planetary boundaries concept - currently being debated, adapted and elaborated – is necessary despite its usefulness. It is important to even briefly acknowledge that the concept of planetary boundaries and the related notion of a 'safe operating space for humanity' are not a purely scientifically determinable but are entangled with social norms, values and processes; therefore a need to democratize the concept has been articulated (Pickering & Persson, 2020). This, along with the question – *What processes are causing us to cross planetary boundaries, and who benefits from these processes?* gives us the opportunity to revisit climate justice and its entanglement with each of the metaconcepts. Currently I am moving toward the phrasing: 'diverse sustainable social-ecological cultures' rather than 'safe operating space for humanity.' I am also in the process of integrating a study (O'Neill et al., 2018) that indicates that "no country meets basic needs for its citizens at a globally sustainable level of resource use," which sobering fact indicts the socio-cultural-economic system of modern industrial civilization.

3. <u>Complexity</u>

Three misconceptions about climate change that I have encountered are –

a) How can humans possibly affect something as large as the Earth?

b) If climate change is really happening, why not wait to fix it until, for example, the weather in Boston is like North Carolina?

c) Technology is going to fix the problem!

Science as we know owes a lot to reductionism, the idea that to understand the whole of a system, we need to divide it into parts and study the parts and their roles. However there are many real world systems that are not fully amenable to reductionism; for these the Aristotelian holistic dictum "the whole is greater than the sum of its parts" applies. Complex systems include the human nervous system, the human endocrine system, ecosystems, social networks, global financial networks and global climate. While working physicists are well aware of the limitations of unbridled reductionism, and often employ what might be called 'optimal reductionism' (Hari Dass, N.D., personal communication, September 15, 2019), the dominance of reductionist methods in science has meant that the detailed study of complex systems as *systems* (rather than solely as a conglomeration of parts) is fairly recent. (It is to be noted that I am not being reductionist about reductionism and holism – we in fact need both). Currently no widely-agreed-upon definition of complexity exists. I align with Paul Fieguth's description of complex systems as large, dynamical non-linear, non-Gaussian, coupled spatial systems (Fieguth, P., 2017). Complex systems are made up of parts that interact non-linearly in time, in ways that make it difficult to predict details of their behavior compared to simple systems. (The division into parts also depends on the question being asked). Some (not all) complex systems exhibit chaotic behavior – for example, weather is chaotic but climate is not (Annan, J., Connolley, W., 2005). Complex systems exhibit hysteresis – that is, having reached a certain state, they cannot be made to return to a previous state simply by reversing the conditions.

Because they are so much easier to study, most systems that students encounter are small, linear and Gaussian. Therefore introducing complexity, especially in the context of climate change in a physics classroom, is challenging.

We present complex systems by first talking about simple systems, with the example of the analog clock. In this case understanding the parts (springs, gears, etc.) leads us to understanding the clock as a whole, because the interactions between parts are simple and clear. The role and significance of the parts do not change as the clock keeps time. Thus the clock may be complicated, but it is not complex.

We then introduce complex systems as systems in which the interactions are as important as the parts (terms like strong coupling and non-linearity can be introduced in higher level physics classes) so that merely understanding the parts and how they function cannot lead us to a classical understanding (including predictability of details) of the whole. An example of a complex system is the human endocrine system, where a knowledge of how each gland works is insufficient to predict the exact behavior of the system as a whole. A lab or demonstration that contrasts the behavior of a simple pendulum with that of the chaotic pendulum is also instructive in making this clear to students. It is important to emphasize that complex systems, while not predictable in the same way that the trajectory of a projectile is predictable, do not behave randomly – we can glean important information from them, but it is not always useful to ask the same questions of a complex system as we do of simple systems.

Once the idea of a complex system is introduced and a couple of examples presented, students often begin to recognize the presence of such systems everywhere around them. This is because most real world systems are complex; the conceptual scaffolding afforded by our brief foray into complexity allows students for the first time to make sense of hitherto unacknowledged everyday presences of complex systems.

In our discussion we point out the ways in which complex systems differ from simple systems. For the purposes of studying climate change, the following two characteristics are important:

a) Feedback loops, both stabilizing (negative) (such as temperature regulation in the human body, or carbon uptake by the solid Earth on timescales of over 100,000 years) and destabilizing (positive) such as the ice-albedo feedback in the climate system; feedback loops exist in simple systems as well, but complex systems may possess multiple feedback loops (many of which could be positive) that can interact with each other at different scales with varying levels of complexity; further, it is possible for feedback loops to flip from negative to positive, as in ocean absorption/ emission of $CO_2$; and

b) Tipping Elements: aspects of the climate system that, once certain thresholds are crossed, result in the system being committed to change in a particular direction, although the timing of that change can vary from abrupt to gradual. This includes for example the melting of the ice sheets. Such changes are generally irreversible on human timescales (Lenton et al., 2008a; Lenton, Timothy M. et al., 2019)

The possibility that interacting positive feedback loops might cause the Earth system as a whole to commit to a systemic change (via a 'tipping cascade') is explored in an important recent paper (Steffen et al., 2018).

The discussion on global climate as a complex system is begun by referring again to Fig. 1 and considering the Earth's sub-systems as one way (among others) of subdividing the Earth into parts. The lines from the sun to the Earth, and those connecting each bubble, represent interactions, which, we remind students, are strong, generally nonlinear, and changing with time (for example by revisiting the fact that small changes in insolation due to orbital variations are connected with changes in ocean currents and atmospheric $CO_2$ levels that together give rise to the dramatic phenomenon of glacial-interglacial cycles.) Students then study graphs and animations of the sea ice extent in the Arctic. The ice is melting because the average surface temperature in the Arctic is rising at twice the global rate. Why might that be happening? The current consensus seems to be rising carbon dioxide levels, but this is a topic of ongoing research, and while the quantification of various feedbacks in the Arctic is an active area of inquiry (Goosse et al., 2018; Stuecker et al., 2018), the Arctic provides some of the most pedagogically useful examples of positive feedback loops in the climate system. We first clarify what positive and negative feedbacks are, through everyday examples. Then, while acknowledging that we don't as yet know to what extent these feedbacks dominate Arctic warming in the current era, we explore the ice-albedo and the permafrost methane feedbacks, which are both positive (in the sense of additive or destabilizing; we use the latter term for clarity). Another destabilizing feedback loop I discuss is one that involves drilling for oil and gas in the Arctic: as the Arctic loses sea ice, the reserves of oil and gas on the sea bed become more accessible, facilitating more drilling operations. As more oil and gas continues to be burned, more sea ice melts, further facilitating Arctic drilling. The two 'natural' feedback loops interact with each other and with the Arctic drilling feedback loop, potentially accelerating warming. This dramatic illustration of nonlinearity helps answer the question of how an initially small effect can become very large very quickly. We also briefly discuss how stabilizing feedback loops (such as ocean absorption of carbon dioxide, or the role of tropical forests in sequestering carbon) can become destabilizing under warmer conditions. The issue of tipping points (or tipping elements) potentially leading to global tipping cascades then becomes clearer – passing several tipping points is likely to lead to irreversible, runaway or catastrophic climate change, after which human intervention is not likely to stabilize or reverse it.

The science of complex systems can be contextualized through the idea of paradigm shifts in the history of science; I find the term 'paradigm' pedagogically useful despite the banality and ubiquity of its current usage. I begin with 'paradigm' in the sense of a 'disciplinary matrix' elucidated in a later work of Thomas Kuhn (Second Thoughts on Paradigms), as a collection of practices, symbolic generalizations, models and exemplars that make and are made by a scientific community (Bird, 2018; Kuhn, Thomas, 2012). I extend this notion of a scientific paradigm to encompass social groups beyond the scientific community which, even if they are not aware of the details of the science, support, benefit from, are affected by and promulgate a worldview built (accurately or not) from this disciplinary matrix so that it becomes the scaffolding for a mostly unquestioned consensus reality. Although there are many critiques of Kuhn's formulation of the progress of science via 'normal science' and 'revolutions,' I find some utility in the sociological-scientific extension of the idea of a paradigm – and paradigm shifts - as formulated by Kuhn. I acknowledge that this extension is an oversimplification of complex historical and socio-cultural processes, disputes and debates; however, inspired by Shapin (Shapin, Steven, 2018), I aver that following certain historical threads in order to understand how we got to this moment in human history can be a worthwhile endeavor. Using the term paradigm in this broader sense, I attempt to trace the development of certain ways of thinking and conceptualizing that are prevalent and dominant today.

In the classroom I motivate this discussion through a brief exploration of the shift from geocentrism to heliocentrism, and the changes wrought in both the cosmological and sociological realms. I follow this with a discussion of ancient to modern conceptualizations of the atom. These topics are already in the syllabus of the course Physics, Nature and Society. I then introduce what may be called the Newtonian paradigm (admittedly somewhat unfair to Newton the person), also referred to as the Mechanistic or 'clockwork universe.' The term 'Newtonian Paradigm' encompasses the mechanical philosophies of Boyle, Kepler, Mersenne, and Descartes among others, and their mathematization via Newton's laws, as well as the atomic materialism of Boyle and Descartes (Boas, 1952; Derek J. de Solla Price, 1964; Shapin, Steven, 2018). This amalgam gives rise to a view of the universe that is reductionist, mechanistic, and deterministic (Heylighen, F. et al., 2007). The analog clock is immediately seen as a metaphor of the Newtonian paradigm. To give students some idea of the historical complexity that accompanies the development of large ideas, we briefly discuss objections to the mechanistic universe (for example William Blake's (Moore, A., 2014)).

We first critique the Newtonian Paradigm from the physics perspective. While Newtonian physics is powerful, we know that it has a limited domain of validity. It fails in the realm of the very small, very fast, and very massive. I introduce the diagram shown (Fig 2). The x-y plane is inspired by very similar diagrams that have appeared in general physics textbooks (French, A.P., 1971; Hobson, Art, 2009). Note that the various conceptual realms do not have hard boundaries, and there is considerable overlap between them; thus Newtonian physics can be considered an approximation of quantum physics, and from the other direction, of relativity. My addition of the z axis acknowledges a revolution in physics and beyond, which is the physics of complexly entangled systems at all scales. The gradation from simple to complex is qualitative, as we don't yet have measures of complexity, and the designation of a system as 'complex' depends in part on the question we are interested in asking, since the same system, when considered from different angles, may be simple or complex. Fig. 2 allows students to understand the damage done when we think within the Newtonian box about systems that exist outside it – for example, some people in the medical profession still refer to the human body as a machine, when in fact the mechanistic model of the body is severely limited (as any endocrinologist or neurologist would agree). Even for a complex nonlinear system like global climate, the mechanistic view can be encountered in popular thinking (for example a 2009 planetarium show about Earth systems and climate change, Dynamic Earth (E&S Shows, 2009), refers to the Earth explicitly as "a machine which is the sum of its parts"). Yet the clockwork, mechanistic analog is a poor one for a system in which the interactions between the parts is so significant that it can change the nature of the parts – where oceans or forests might become, for example, net emitters rather than absorbers of carbon.

    The power of this diagram is that it makes the invisible visible – our conceptual frameworks through which we make sense of the world suddenly become apparent, and unexamined or default assumptions become susceptible to scrutiny. Classroom debates and discussions based on this diagram often center on the possibility that science itself changes and evolves, and that a paradigm shift (in the broader sense) may well be taking place at this moment in history. The extent to which old ways of thinking persist despite changes in scientific understanding is also an interesting topic of discussion that complicates the idea of a paradigm shift. For instance we may point to influences of Einstein's theory of relativity in philosophy and art, but the way humans live and think in modern industrial societies is mostly unaffected (except for the use of technological innovations owed to relativity) by this seismic shift in

theoretical physics. There are many factors that prevent scientific revolutions from changing the way people at large conceptualize the world but among these is the role of power structures in society. Who benefits from the status quo? How do systems of power control how people conceptualize the world? While a physics class is not adequate for answering such questions in any depth, simply raising them makes for interesting speculative discussions, and prepares us for thinking about climate justice. Students also look at the pros and cons of large scale technofixes (geoengineering schemes such as the injection of aerosols in the upper atmosphere to reflect away sunlight) (Oxford Geoengineering Programme, 2018) via this metacognitive view of science – are some geoengineering schemes an example of Newtonian thinking about a non-Newtonian problem? I introduce the term *paradigm blindness* – the notion that a (socio-cultural or scientific) paradigm in which one is immersed makes it extremely difficult to realize, understand, or consider seriously an alternative worldview that has a different underlying conceptual structure. The realization that a genuinely sustainable world involves not only technological and societal change but also a major worldview shift is revelatory for many students.

The idea that complex systems science is an essential component of effective climate change pedagogy is supported by other researchers (Roychoudhury, A. et al., 2017). My experience indicates that enlarging the discussion of complexity beyond physics to the historical and sociological context enriches student understanding. To take full advantage of these conceptual revelations, it is important that the teaching practice incorporate systems thinking methodology throughout - for example I use concept maps extensively, as well as small-group work, stock-and-flow analogies such as the Carbon Bathtub, and discussion of real-world dilemmas and situations. However, far more work needs to be done to develop a truly systems-based pedagogy in the context of climate change.

**Interdisciplinarity: Climate Impacts and Climate Justice**

The impacts of climate change on human societies and ecosystems are discussed during Climate Week, where class time and lab time are devoted to a deeper exploration of the uneven impacts of climate change on societies. Students explore sea level rise, weather 'weirding,' extreme weather, including potentially fatal heat waves and 'unliveability' in some areas, large scale human migration, mass extinctions of species, rising food and water insecurity, increasing evidence connecting biodiversity loss with the emergence of new zoonotic diseases, and the poleward movement of tropical diseases. Higher level general physics courses for science majors do not allow sufficient time for a deep exploration of the societal consequences of climatic impacts. Nevertheless it is important for ethical reasons to devote some time to making these connections, consolidating the earlier discussions of climate justice, and thereby framing climate change at the nexus of socio-economic, historical, political and environmental concerns as an issue of justice. Further, no meaningful discussion of solutions can be undertaken without such a framework, as we will see in the next section. In classes with more flexibility, such as my physics course for non-science majors (Physics, Nature and Society), a subset of topics within this frame can be explored in more detail.

I generally initiate the discussion on interdisciplinarity by asking students how many disciplines they think are involved in considering the impacts of climate change. Students come up with Biology, Economics, Sociology, (for example) and we expand the list beyond these by building a web of disciplines on the whiteboard, with key questions that pertain to climate change. For example under 'Sociology' we could ask 'What populations are disproportionately affected by climate change?' This exercise gives students the range of disciplines that are involved in engaging with the problem of climate impacts. They can then locate their majors and desired career paths within these disciplines, which helps make climate change relevant to their lives.

Climatic impacts, as we have mentioned, are not equally distributed; the rich are likely to ride out some of the worst effects of climate change (Bendix, A., 2019; UN News, 2019). The people most affected by climate change – populations of the global South, people of color, the economically disadvantaged, indigenous people and the young – are also people who have contributed least to creating the problem, hence the centrality of climate justice. This is why Fig. 1 does not represent the 'human' impact on Earth's systems through the term 'Anthroposphere' but distinguishes between modern industrial civilization and those humans whose impact is low to nonexistent. Those most affected by climate impacts often have less access to resources and infrastructure. It is important to point out that although the prevailing economic and political power structures put certain groups of people at a disadvantage, these groups are not necessarily helpless bystanders to their misfortunes. For example, indigenous people, despite being only 5% of the world's population, manage about 80% of the world's biodiversity, helping to keep nearly 300,000 million tons of carbon from the atmosphere (Rights and Resources Institute, et al., 2018). Multiple reports (Etchart, 2017; Reytar, K. & Veit, P., 2016) indicate that indigenous people, who are often at the frontlines of resistance to extractive industries, are essential for protecting the world from environmental crises, including climate change. The US youth suing the federal government for not protecting them from climate change (Schwartz, J., 2020) and the more recent international youth movements provide fuel for the discussion of the role of the young in securing a sustainable future (Tigue, K., 2019). There is now wide recognition that for a sustainable future we must also solve the problem of social inequality (IPCC, 2014).

Depending on the class and time available, we explore selected topics on climate justice. My approach is not to present pre-packaged or partisan answers, but to encourage students to explore in groups and present to the class from a wide range of resources. For example one group might compare the carbon emissions of nations, total and per capita, both historically and in the present day, while another group could study different climate justice movements, and a third group could take up the impact of climate change on various marginalized groups within nations. Each group would then 'teach the class.' Studying climate justice movements is especially popular with students because they see that others are concerned and are taking action. A more transdisciplinary approach is to discuss news items that foreground climate justice in the context of current events – for example I have often used the excellent account by John D. Sutter of the tragic death of an African-American woman, Stacy Ruffin, during the Louisiana floods of 2016, which combines a moving account of a family tragedy with discussions of racial and economic injustice and climate science (Sutter, John D., 2017). These approaches make the climate issue feel relevant and urgent, and may turn potentially disengaged non-science majors 'on' to the subject, as they see the connection between climate change, their lives and their majors.

Climate justice is also relevant in our discussion of climate action and solutions.

**Interdisciplinarity: Climate Action - Barriers and Solutions through the Meta-Conceptual Framework**

A common student response to learning about climate change is "What can be done about it?" Students get climate solutions information from social media and are not necessarily able to judge the efficacy of these solutions. Individual-action 'solutions' such as changing the lightbulbs or recycling or keeping your tires inflated are often amplified in the media; although these may be helpful if scaled up, they do not go to the heart of the climate problem. Students often sense that the problem is much bigger than their individual selves, as evidenced by comments such as "what can I do, I'm just one person." This is a possible additional reason why learning about climate change results in despair. In a state university where students are largely first generation working class, it is possible that balancing work and school, family responsibilities and perhaps a general sense of disempowerment makes it difficult for students to think that any action of theirs can 'fix the climate.' Thus students tend to take refuge in the possibility that technology will provide the solution, and that someone else - government, corporations, scientists - will take care of the problem, although the continuing lack of action on climate change and worsening climatic conditions create discomfort with that assumption.

Therefore it is useful to introduce climate solutions through a framework that we describe below.

First, our discussion of climate science, along with the key conclusions of the IPCC Special Report (2017), leads us to the following statement:

*To restore the climate, the carbon cycle must be brought back into balance (steady state) such that atmospheric $CO_2$ levels are commensurate with a rise in average global surface temperature of no more than 1.5 ° C compared to pre-industrial times.* Whether and how this is possible becomes a focus of a lively discussion. Among the first potential solutions is of course

switching immediately to green power – solar, wind, etc. and putting an end to all mining and burning of fossil fuels.

However, green power also requires extraction, and a recent report states that about 50% of greenhouse gas emissions come from mining and other extractive processes, which are also responsible for 90% of biodiversity loss and water stress (UNEP, 2019). So simply switching the energy source without changing the system in profound ways is unlikely to be a long-lasting solution.

We then broaden our discussion of climate change by reminding ourselves of the metaconcept of Planetary Boundaries. To the statement above, we add:

*Other Planetary Boundaries must be respected in order for sustainable human civilizations to flourish.*

The above reminds us of the importance of other planetary boundaries, such as the anthropogenic impact on the nitrogen cycle and the massive threat to biodiversity; we can take this opportunity to relate climate change to a wider set of complex problems.

These discussions involve challenging and transcending various dichotomies that are made apparent by the climate crisis. These include: Individual/ Society, Local/ Global, Human/ (Rest of) Nature and Economics/ Environment. So for example we can emphasize that while individual solutions may be meaningful personally, they will not have an appreciable effect on the problem without being multiplied many times over. So how might one bridge the gap between individual action and a systemic shift? The Local/ Global dichotomy is similar. Could some useful local action – for example, having local government provide public transport – be made easily adaptable across many towns through a creative use of the internet?

The apparent dichotomy between economic needs and environmental protection leads us to a discussion of the current model of development and economic theory. This also follows logically from our brief acknowledgment of the violation of other planetary boundaries, which emphasize that we cannot look at climate change in isolation from biodiversity loss or the imbalance in the nitrogen cycle or land system change. All of these problems have their roots in the socio-economic paradigm that scaffolds modern industrial civilization. While we cannot do full justice to these topics in the physics classroom, it is worth pointing out that the basic assumptions of classical mainstream economics are very much consistent with the Newtonian paradigm – the focus on the fundamental unit of society being a rational individual human acting in self-interest is a kind of social atomism. An examination of the history of economics (Ackerman, F., 2018) reveals a fascinating tendency of early economists to try to model economic behavior inappropriately on concepts from physics. The notion of endless growth so beloved of mainstream economists has been challenged by physicists and others as being incommensurate with natural laws and other limits (for instance see (Dhara, C. & Author, 2021)). Time permitting, the connection between greenhouse gas emissions and GDP can be introduced, with a discussion on the exponential growth of these along with material resource use (Wiedmann et al., 2015). The runaway greenhouse-gas-emitting machine that is modern industrial civilization, when held against natural limits and other constraints, including planetary boundaries, points to the failure of mainstream economics to acknowledge physical reality. Alternative economic models that are conscious of planetary boundaries include Oxford University economist Kate Raworth's 'Doughnut economics' – her TED talk on this generates a lively discussion in the classroom (Raworth, K., 2018). Our brief earlier excursion into indigenous epistemologies connects with the need for a paradigm shift already discussed in the course in the context of the Newtonian Paradigm.

It is clear at this point to students that we cannot usefully discuss climate solutions without going on necessary multidisciplinary excursions outside physics.

When we examine proposed solutions via the metaconcepts framework, we ask – *Does the solution aim to restore balance? Does it restore other planetary boundaries, the maintenance of which is also key to our survival? Does it take complexity into account? What's the underlying paradigm? Is the solution imagined in a systems way? Is it effective systemically? Is it just and equitable to all concerned?* Consider replacing gasoline-powered cars with electric cars. This is of course going to reduce $CO_2$ emissions over the lifetime of the car. But what about the disturbance to the carbon cycle involved in mining and manufacture? And what if the car is charged via power that comes from burning fossil fuels? What about the lithium for the batteries? Considering that serious consideration of mining the moon for minerals is motivated in part by the green energy revolution (McLeod, C.L. & Krekeler, M.P.S., 2017) what is the ultimate carbon footprint of 'green' cars? What about public transport becoming more available instead of everyone owning a car? How about redesigning cities and work such that people don't have to commute very far? Can we use new (and old) ideas from architecture that reduce or eliminate the need for air-conditioning? Through such discussions and imaginative exercises ('design and sketch a future city') and occasionally reading science fiction stories, students recognize that long-lasting and equitable climate solutions cannot be purely technological, but must entangle with socio-economic changes that respect planetary boundaries and creatively occupy the 'safe operating zone' of diverse, sustainable, equitable human cultures. Technological innovations must minimize the climate impact over the entire life cycle of the product, and additionally must respect *all* planetary boundaries. This leads to the inescapable conclusion that true sustainability can only arise if we flatten all resource use curves (Dhara, C. & Author., 2021). The fact that no nation currently meets human needs within planetary boundaries (O'Neill et al., 2018) calls for a radical reimagining of human socio-economic and human-Nature relationships. Central to these are considerations of equity and justice, which increase community resilience to climate impacts and allow those who are most impacted to benefit from and contribute to the mitigation of the crisis.

- Thus when discussing purported climate solutions, whether these be the transition to green energy, or geoengineering schemes such as Solar Radiation management, we ask the questions: Will the solution move us toward restoration of the carbon cycle? Will it move us within other planetary boundaries (such as the restoration of the Nitrogen cycle, and of biodiversity)?
- Is the solution just and equitable? What are its underlying values? Does it move the most marginalized groups in society toward increased resilience to climatic and related impacts? Does it allow these groups to participate in contributing to climate mitigation? Is it amenable to the thriving of our ecosystems?
- What is the paradigm from which the solution arises? Does it take into account the complexity of natural-cultural systems?

However, it is one thing to propose solutions, and quite another to act upon them. Some of the items listed under "Challenges to Teaching Climate Change" are also barriers to action. We discuss in particular epistemological, cultural and psychological barriers, and add one more: power and vested interests. The disproportionate influence of the fossil fuel industry on the economy and politics makes it extremely unlikely that those who benefit from the sociological power structure are going to change or dismantle it. The paper trail of this influence, monetary and otherwise, has been revealed via investigative journalism and scholarship (Brulle, 2014).

Because we are dealing with a vulnerable population – young students – the psychological dimension of learning about climate change must be addressed. This is best done within the framework of transformational learning, where the classroom culture allows for a sense of community building and each student feels valued. The dire nature of the climate crisis inevitably leads many students toward despair and hopelessness. It is important not to gloss over this reaction, to trivialize or ignore it. Instead we discuss climate scientist Steve Running's idea of the 5 stages of 'climate grief,' and variations thereof (based on the Kubler-Ross scale for stages of grief) (Running, 2007). During the course of the semester I invite students to share anonymously (or openly, if they are so inclined) where they think they might be on this scale at the moment, and whether they think it is a useful measure of the psychological response to learning about climate change. This allows students to accept what they might be feeling – despair, anger, apathy – as normal reactions that can transition to grieving and accepting the reality of climate change – and ultimately, we hope, harnessing the emotional energy to becoming changemakers (McMakin, D. & Author, 2021). I have on four occasions invited a professor of environmental poetry to my class to see whether creative writing can help students navigate grief and other difficult emotions. In classes for non-science majors I find that students who would not normally engage in discussing science suddenly begin to see its relevance to the real, complex world we inhabit, and begin to take an interest.

One additional factor in increasing student involvement and interest is to create opportunities for going beyond the classroom. For example one physics class for non-science majors wrote a comprehensive constructive critique of a climate change exhibit at a large local science museum that resulted in an appreciative letter of acknowledgment from the museum. On another occasion, non-science majors presented the science of climate change to an Expository Writing class that was studying it from a literary angle. Students of Physics, Nature and Society taught climate science basics to a senior poetry class, following which the poetry students led a poetry workshop centered around the complex emotions of grief, anger, hope and loss that were evoked. Another class participated in a national traveling exhibition on the American Family by creating a display about Hurricane Maria's impact on the people of Puerto Rico. More recently, a freshman seminar class on climate change created a Heat Impact Awareness Tool for the elderly and presented it to local city health officials. These efforts to bring climate change outside the classroom help students feel they are doing something meaningful with what they've learned.

**Results and Conclusion**

In surveys, students self-report a significant leap in understanding the basic physics of climate change by the end of the course. Homework assignments, exams and tests support this – students rarely, if ever, reproduce common misconceptions (such as confusing climate change with the ozone problem), and are able to explain the basics of the carbon cycle, greenhouse effect, and the role of complexity in the climate system. The use of holistic visual tools such as Fig. 1, integrated with the three metaconcepts, initially challenges students, as they are not used to making connections across subtopics and disciplines. But with repeated use every time a climate topic is discussed, preliminary results suggest that students are better able to use the tool for a holistic understanding by the end of the semester – for example, I observe greater fluency and comfort with using Fig. 1 to explain the basics of the anthropogenic impact on the carbon cycle.

With regard to our explorations of justice issues and the epistemological and scientific critiques of the socio-economic norms that form the foundations of modern industrial civilization – classroom discussions as well as formal and informal feedback provide evidence of greater engagement and interest among students. In the non-science majors course Physics, Nature and Society, test and exam questions include these relevant excursions outside physics in the form of short-answer and short-essay questions. Along with end-of-semester surveys, these results indicate that the vast majority of students (80-90% per class) over the years appreciate the deeper insights that our interdisciplinary explorations bring to their understanding of climate change. Students are, for example, able to articulate how climate justice intersects with the concerns and activism of marginalized groups such as indigenous peoples.

Despite these encouraging signs, it is important to note that this work is still at a preliminary stage. As mentioned before, the small N (about 60 students total in physics courses for science majors, and about 50 in courses for non-science majors total over a period of five years) and the incremental nature of the approach described in this paper (see 'A Note on Teaching Approach and Method') does not allow us to generalize, and instead points to the need for further study, quantitative as well as qualitative.

A major disadvantage of my approach, and indeed of any approach limited to one course or classroom, is that students' interdisciplinary understanding of climate change, and any progress toward an epistemic shift is not reinforced in other classes, and is thus subject to attrition over time. This highlights the need for a cross-disciplinary collaborative framework at the school or university level in which students learn about climate change from multiple disciplinary perspectives in an integrated way, based on transdisciplinarity and transformational learning. It is possible that the framework described in this paper can be adapted for such a cross-disciplinary undertaking, but, as mentioned earlier, the success of such a framework depends critically on the teaching approach, and therefore faculty training in both pedagogy and content, in which faculty themselves experience the power of transformational learning through epistemic shifts, is crucial. The inter-/transdisciplinary nature of the climate problem is indeed a significant stumbling block. As described by scholars of transdisciplinarity (Brown, V. A. et al., 2010; Leavy, 2011) there are many challenges to going 'beyond the discipline,' not least among which is the fact that each discipline has its own practices and paradigms that are the default comfort zones of the practitioners. In 2017 I was part of a team of three that co-conceptualized and co-ran a summer workshop (Wade Institute for Science Education, 2017) for middle and high school science teachers in the US Northeast, focused on an interdisciplinary approach to teaching climate change, based in part on an earlier version of my framework. While the teachers were enthusiastic and eager to learn across the disciplines, they found the interdisciplinary nature of the experience most challenging, and most difficult to integrate into projects for their students. This is not surprising, given that conventional education tends to compartmentalize the imagination into disciplinary boxes with nearly impenetrable walls. Thus faculty training in the relatively new field of inter-/ transdisciplinary pedagogy is necessary.

So far our framework is more interdisciplinary than transdisciplinary, although it shares some of the features of transdisciplinary learning (Brown, V. A. et al., 2010; Leavy, 2011). A transdisciplinary climate pedagogy framework need not develop incrementally from an interdisciplinary one. Well regarded approaches such as Project-Based Learning, Problem-Based Learning, Case-Study- Based learning and Social Emotional Learning (English & Kitsantas, 2013; Weissberg et al., 2015; Yale Poorvu Center for Teaching and Learning, n.d.) can contribute to the development of a framework that is transdisciplinary from the get-go.

One major drawback of a conventional approach to teaching climate science is that the climate crisis is presented in isolation from social-economic-political forces as well as other major environmental and social issues.  This enables the kind of blindness that is evident in mainstream approaches to the climate problem, including discourses at the level of the UN and other elite bodies, where an almost exclusive focus on the carbon cycle encourages a narrow technological approach to 'solutions.' This ignores other violations of planetary boundaries such as biodiversity loss, the disruption of the nitrogen cycle, and massive land system changes. All of these are grounded in an economic model based on endless growth and consumption, to the detriment of large masses of people and nonhuman species. These are generally unchallenged in the mainstream climate discourse, thereby leading to deep contradictions and divergences between intent, policy and practice, such as is evident in the UN's Sustainable Development Goals (Dhara, C. & Author, 2021). No 'climate fix' is possible without looking at these problems in a connected way through a transdisciplinary, justice-based, epistemological lens. In its present form, my framework does not go far enough in integrating these related issues. Ideally the study of climate science should be a gateway toward the realization that climate is only one of a number of interrelated issues in which other grave environmental problems are entangled with socio-economic structures and histories. My current framework is only partially successful in this regard. I am working toward a more holistic integration of these ideas for the next version of my framework.

As mentioned in A Note on Teaching Approach and Method, the primary purpose of this paper is to make a case for inter-/transdisciplinarity and transformational learning in teaching climate change in a science classroom. I do not claim that this framework is (in its current form) more than partially successful, or that it is the only possible framework, or that it is appropriate for every context and setting. If this paper inspires fellow educators to critique conventional thinking on climate/ environmental pedagogy, and through their own experiments develop a tapestry of diverse inter-/transdisciplinary frameworks suited to their contexts, then it will have accomplished a large part of its purpose.

**References**

Ackerman, F. (2018). *Worst-Case Economics: Extreme Events in Climate and Finance*. Anthem Press.

Annan, J., Connolley, W. (2005, September). Chaos and Climate. *RealClimate*. http://www.realclimate.org/index.php/archives/2005/11/chaos-and-climate/

APA Taskforce on Interface Between Psychology and Global Climate Change. (2009). *Psychology and Global Climate Change: Addressing a Multifaceted Phenomenon and Set of Challenges*. American Psychological Association.

Archer, D. (2010). *The Global Carbon Cycle*. Princeton University Press.

Archer, D., Eby, M., Brovkin, V., Ridgwell, A., Cao, L., Mikolajewicz, U., Caldeira, K., Matsumoto, K., Munhoven, G., Montenegro, A., & Tokos, K. (2009). Atmospheric Lifetime of Fossil Fuel Carbon Dioxide. *Annual Review of Earth and Planetary Sciences*, *37*(1), 117–134. https://doi.org/10.1146/annurev.earth.031208.100206

Bain, K. (2004). What Makes Great Teachers Great? *The Chronicle of Higher Education: The Chronicle Review*, *50*(31), B7.

Bain, K., & Zimmerman, J. (2009). *Understanding Great Teaching* [Text]. Association of American Colleges & Universities. https://www.aacu.org/publications-research/periodicals/understanding-great-teaching

Ballew, M. T., Leiserowitz, A., Roser-Renouf, C., Rosenthal, S. A., Kotcher, J. E., Marlon, J. R., Lyon, E., Goldberg, M. H., & Maibach, E. W. (2019). Climate Change in the American Mind: Data, Tools, and Trends. *Environment: Science and Policy for Sustainable Development*, *61*(3), 4–18. https://doi.org/10.1080/00139157.2019.1589300

Barad, K. (2007). *Meeting the Universe Halfway: Quantum Physics and the Entanglement of Matter and Meaning*. Duke University Press.

Barnhardt, R., & Kawagley, A.O. (2005). Indigenous Knowledge Systems/ Alaska Native Ways of Knowing. *Anthropology and Education Quarterly*, *36*(1), 8–23.

Bell, D. A., & Marcum, J. C. (2018). Adapting Three Classic Demonstrations To Teach Radiant Energy Trapping and Transfer As Related to the Greenhouse Effect. *Journal of Chemical Education*, *95*(4), 611–614. https://doi.org/10.1021/acs.jchemed.7b00626

Bendix, A. (2019, June 10). 45 Unreal Photos of Billionaire Bunkers That Could Shelter the Superrich During an apocalypse. *Businessinsider.Com*. https://www.businessinsider.in/science/45-unreal-photos-of-billionaire-bunkers-that-could-shelter-the-superrich-during-an-apocalypse/articleshow/69729435.cms

Bird, A. (2018). Thomas Kuhn. In E. N. Zalta (Ed.), *The Stanford Encyclopedia of Philosophy* (Winter 2018). Metaphysics Research Lab, Stanford University. https://plato.stanford.edu/archives/win2018/entrieshomas-kuhn/

Boas, M. (1952). The Establishment of the Mechanical Philosophy. *Osiris*, *10*, 412–541. https://doi.org/10.1086/368562

Boström, M., Andersson, E., Berg, M., Gustafsson, K., Gustavsson, E., Hysing, E., Lidskog, R., Löfmarck, E., Ojala, M., Olsson, J., Singleton, B. E., Svenberg, S., Uggla, Y., & Öhman, J. (2018). Conditions for Transformative Learning for Sustainable Development: A

Theoretical Review and Approach. *Sustainability*, *10*(12), 4479. https://doi.org/10.3390/su10124479

Brown, V. A., Harris, J., & Russell, J. (2010). *Tackling Wicked Problems Through the Transdisciplinary Imagination*. Routledge.

Brulle, R. J. (2014). Institutionalizing delay: Foundation funding and the creation of U.S. climate change counter-movement organizations. *Climatic Change*, *122*(4), 681–694. https://doi.org/10.1007/s10584-013-1018-7

Carbon Tracker Initiative. (2017, August). *Carbon Bubble*. Carbon Tracker Initiative. https://carbontracker.org/terms/carbon-bubble/

Chess, C., & Johnson, B. B. (2007). Information is not enough. In L. Dilling & S. C. Moser (Eds.), *Creating a Climate for Change: Communicating Climate Change and Facilitating Social Change* (pp. 223–234). Cambridge University Press; Cambridge Core. https://doi.org/10.1017/CBO9780511535871.017

Cook, J., Oreskes, N., Doran, P. T., Anderegg, W. R. L., Verheggen, B., Maibach, E. W., Carlton, J. S., Lewandowsky, S., Skuce, A. G., Green, S. A., Nuccitelli, D., Jacobs, P., Richardson, M., Winkler, B., Painting, R., & Rice, K. (2016). Consensus on consensus: A synthesis of consensus estimates on human-caused global warming. *Environmental Research Letters*, *11*(4), 048002. https://doi.org/10.1088/1748-9326/11/4/048002

Cook, J., Supran, G., Lewandowsky, S., Oreskes, N., & Maibach, E. (2019). *America Misled: How the fossil fuel industry deliberately misled Americans about climate change*. George Mason University Center for Climate Change Communication.
54

Derek J. de Solla Price. (1964). Automata and the Origins of Mechanism and Mechanistic Philosophy. *Technology and Culture*, *5*(1), 9–23. JSTOR. https://doi.org/10.2307/3101119

Dhara, C., & Author. (2021). Forthcoming.

Dweck, C.S. (2006). *Mindset: The New Psychology of Success*. Random House.

English, M., & Kitsantas, A. (2013). Supporting Student Self-Regulated Learning in Problem- and Project-Based Learning. *Interdisciplinary Journal of Problem-Based Learning*, *7*(2). https://doi.org/10.7771/1541-5015.1339

EPA (Archived). (2014). *Causes of Climate Change*. Climate Change Science. https://archive.epa.gov/epa/climate-change-science/causes-climate-change.html

E&S Shows. (2009). *Dynamic Earth—E&S Digital Theater Show*. http://es.com/Shows/dynamicearth

Etchart, L. (2017). The role of indigenous peoples in combating climate change. *Palgrave Communications*, *3*(1), 17085. https://doi.org/10.1057/palcomms.2017.85

Euler, E., Rådahl, E., & Gregorcic, B. (2019). Embodiment in physics learning: A social-semiotic look. *Physical Review Physics Education Research*, *15*(1), 010134. https://doi.org/10.1103/PhysRevPhysEducRes.15.010134

Fieguth, P. (2017). *An Introduction to Complex Systems: Society, Ecology and Nonlinear Dynamics*. Springer.

French, A.P. (1971). *Newtonian Mechanics* (First). W.W. Norton.

Gao, Y., Gao, X., & Zhang, X. (2017). The 2 °C Global Temperature Target and the Evolution of the Long-Term Goal of Addressing Climate Change—From the United Nations
55

Framework Convention on Climate Change to the Paris Agreement. *Engineering*, *3*(2), 272–278. https://doi.org/10.1016/J.ENG.2017.01.022

Goosse, H., Kay, J. E., Armour, K. C., Bodas-Salcedo, A., Chepfer, H., Docquier, D., Jonko, A., Kushner, P. J., Lecomte, O., Massonnet, F., Park, H.-S., Pithan, F., Svensson, G., & Vancoppenolle, M. (2018). Quantifying climate feedbacks in polar regions. *Nature Communications*, *9*(1), 1919. https://doi.org/10.1038/s41467-018-04173-0

Grant, A., & Coffin, M. (2019). *Breaking the Habit: Why None of the Large Oil Companies are "Paris Aligned" and what they need to do to get there*. Carbon Tracker Initiative. https://carbontracker.org/reports/breaking-the-habit/

Hall, Brendan M. (2011). Threshold Concepts and Troublesome Knowledge: Towards a "Pedagogy of Climate Change?" In Haslett, S.K., Gedye, S., & France, D. (Eds.), *Pedagogy of Climate Change*. Higher Education Academy, Geography, Earth and Environmental Sciences.

Hari Dass, N.D. (2019, September 15). *My Climate Pedagogy Article* [Personal communication].

Heylighen, F., Cilliers, P., & Gershenson, C. (2007). Complexity and Philosophy. In Bogg, Jan & Geyer, Robert (Eds.), *Complexity, Science and Society* (First, p. 184). Routledge. https://arxiv.org/ftp/cs/papers/0604/0604072.pdf

Hobson, Art. (2009). *Physics: Concepts and Connections* (Fifth). Pearson.

Hoggan, C.D. (2018). The Current State of Transformative Learning Theory: A Metatheory. *Phronesis*, *7*(3), 18–25. https://doi.org/10.7202/1054405ar

Intergovernmental Science-Policy Platform on Biodiversity and Ecosystem Services. (2019). *Global Assessment Report on Biodiversity and Ecosystem Services*. IPBES Secretariat.

IPCC. (2014). *IPCC, 2014: Climate Change 2014: Synthesis Report. Contribution of Working Groups I, II and III to the Fifth Assessment Report of the Intergovernmental Panel on Climate Change* (p. 151). IPCC.

IPCC. (2018). *Global Warming of 1.5°C: An IPCC Special Report on the impacts of global warming of 1.5°C above pre-industrial levels and related global greenhouse gas emission pathways, in the context of strengthening the global response to the threat of climate change, sustainable development, and efforts to eradicate poverty*. https://www.ipcc.ch/sr15/

Iyengar, R. (2020, April 23). Charting a Roadmap for Radical, Transformative Change in the Midst of Climate Breakdown. *State of the Planet: Earth Institute, Columbia University*. https://blogs.ei.columbia.edu/2020/04/23/workshop-change-climate-breakdown/

Kothari, A., Salleh, A., Escobar, A., Demaria, F., & Acosta, A. (Eds.). (2019). *Pluriverse: A Post-Development Dictionary*. Tulika Books.

Kuhn, Thomas. (2012). *The Structure of Scientific Revolutions* (Fourth). University of Chicago Press.

Kwauk, C. (2020). *Roadmaps to Quality Education in a Time of Climate Change* [Brief]. Brookings Institute. https://www.brookings.edu/wp-content/uploads/2020/02/Roadblocks-to-quality-education-in-a-time-of-climate-change-FINAL.pdf

Lazowski, R. A., & Hulleman, C. S. (2016). Motivation Interventions in Education: A Meta-Analytic Review. *Review of Educational Research*, *86*(2), 602–640. https://doi.org/10.3102/0034654315617832

Leaton, J. (2011). *Unburnable Carbon: Are the World's Financial Markets Carrying a Carbon Bubble?* Carbon Tracker Initiative.

Leavy. (2011). *Essentials of Transdisciplinary Research* (1st ed.). Routledge.

Lee, T. M., Markowitz, E. M., Howe, P. D., Ko, C.-Y., & Leiserowitz, A. A. (2015). Predictors of public climate change awareness and risk perception around the world. *Nature Climate Change*, *5*(11), 1014–1020. https://doi.org/10.1038/nclimate2728

Leiserowitz, A., Smith, N., & Marlon, J.R. (2010). *Americans' Knowledge of Climate Change* (Yale Project on Climate Change Communication, p. 60). Yale University. https://environment.yale.edu/climate-communication-OFF/files/ClimateChangeKnowledge2010.pdf

Lenton, T. M., Held, H., Kriegler, E., Hall, J. W., Lucht, W., Rahmstorf, S., & Schellnhuber, H. J. (2008a). Tipping elements in the Earth's climate system. *Proceedings of the National Academy of Sciences*, *105*(6), 1786–1793. https://doi.org/10.1073/pnas.0705414105

Lenton, T. M., Held, H., Kriegler, E., Hall, J. W., Lucht, W., Rahmstorf, S., & Schellnhuber, H. J. (2008b). Tipping elements in the Earth's climate system. *Proceedings of the National Academy of Sciences of the United States of America*, *105*(6), 1786–1793. https://doi.org/10.1073/pnas.0705414105

Lenton, Timothy M., Rockström, Johan, Gaffney, Owen, Rahmstorf, Stefan, Richardson, Katherine, Steffen, Will, & Schellnhuber, Hans J. (2019). Climate Tipping Points—Too Risky to Bet Against. *Nature*, *575*, 592–595. https://doi.org/doi: 10.1038/d41586-019-03595-0

Levy, B. S., & Patz, J. A. (2015). Climate Change, Human Rights, and Social Justice. *Annals of Global Health*, *81*(3), 310–322. https://doi.org/10.1016/j.aogh.2015.08.008

Lewis, S., & Gallant, A. (2013, August 22). *In Science, the Only Certainty is Uncertainty*. The Conversation. https://theconversation.com/in-science-the-only-certainty-is-uncertainty-17180

Lindsey, R. (2020, February). *Climate Change: Atmospheric Carbon Dioxide*. Climate.Gov: Science and Information for a Climate-Smart Nation. https://www.climate.gov/news-features/understanding-climate/climate-change-atmospheric-carbon-dioxide

Lotz-Sisitka, H., Wals, A. E., Kronlid, D., & McGarry, D. (2015). Transformative, transgressive social learning: Rethinking higher education pedagogy in times of systemic global dysfunction. *Current Opinion in Environmental Sustainability*, *16*, 73–80. https://doi.org/10.1016/j.cosust.2015.07.018

Macintyre, T., Lotz-Sisitka, H., Wals, A., Vogel, C., & Tassone, V. (2018). Towards transformative social learning on the path to 1.5 degrees. *Current Opinion in Environmental Sustainability*, *31*, 80–87. https://doi.org/10.1016/j.cosust.2017.12.003

McLeod, C.L., & Krekeler, M.P.S. (2017). Sources of Extraterrestrial Rare Earth Elements: To the Moon and Beyond. *Resources*, *6*(40). https://doi.org/10.3390/resources6030040

McMakin, D., & Author, V. (2021). Manuscript under Preparation.

Meinshausen, M., Meinshausen, N., Hare, W., Raper, S. C. B., Frieler, K., Knutti, R., Frame, D. J., & Allen, M. R. (2009). Greenhouse-gas emission targets for limiting global warming to 2 °C. *Nature*, *458*(7242), 1158–1162. https://doi.org/10.1038/nature08017

Meyer, J., & Land, R. (2003). *Threshold Concepts and troublesome Knowledge: Linkages to Ways of Thinking and Practicing Within the Disciplines* (Enhancing Teaching-Learning Environments in Undergraduate Courses). School of Education, University of Edinburgh.

Mezirow, J., & Taylor, E.W. (Eds.). (2009). *Tranformative Learning in Practice: Insights from Community, Workplace, and Higher Education*. Wiley.

Moore, A. (2014, December). *Alan Moore on William Blake's contempt for Newton | Blog | Royal Academy of Arts*. https://www.royalacademy.org.uk/article/william-blake-isaac-newton-ashmolean-oxford

Murphy, J. (2009). *Environment and Imperialism: Why Colonialism Still Matters* (No. 20; SRI Papers). Sustainability Research Institute, School of Earth and Environment, University of Leeds.

Odell, V., Molthan-Hill, P., Martin, S., & Sterling, S. (2020). Transformative Education to Address All Sustainable Development Goals. In W. Leal Filho, A. M. Azul, L. Brandli, P. G. Özuyar, & T. Wall (Eds.), *Quality Education: Vol. Encyclopedia of the UN Sustainable Development Goals* (pp. 905–916). Springer International Publishing. https://doi.org/10.1007/978-3-319-95870-5_106

O'Neill, D. W., Fanning, A. L., Lamb, W. F., & Steinberger, J. K. (2018). A good life for all within planetary boundaries. *Nature Sustainability*, *1*(2), 88–95. https://doi.org/10.1038/s41893-018-0021-4

Oxford Geoengineering Programme. (2018). *What is Geoengineering?* http://www.geoengineering.ox.ac.uk/www.geoengineering.ox.ac.uk/what-is-geoengineering/what-is-geoengineering/index.html

Pendrill, F., Persson, U. M., Godar, J., Kastner, T., Moran, D., Schmidt, S., & Wood, R. (2019). Agricultural and forestry trade drives large share of tropical deforestation emissions. *Global Environmental Change*, *56*, 1–10. https://doi.org/10.1016/j.gloenvcha.2019.03.002
60

Pickering, J., & Persson, Å. (2020). Democratising planetary boundaries: Experts, social values and deliberative risk evaluation in Earth system governance. *Journal of Environmental Policy & Planning*, *22*(1), 59–71. https://doi.org/10.1080/1523908X.2019.1661233

Plutzer, E., McCaffrey, M., Hannah, A. L., Rosenau, J., Berbeco, M., & Reid, A. H. (2016). Climate confusion among U.S. teachers. *Science*, *351*(6274), 664–665. https://doi.org/10.1126/science.aab3907

Raworth, K. (2018, April). *A Healthy Economy Should Be Designed to Thrive, Not Grow*. https://www.ted.com/talks/kate_raworth_a_healthy_economy_should_be_designed_to_thrive_not_grow?language=en

Reytar, K., & Veit, P. (2016, November). Indigenous People and Local Communities Are the World's Secret Weapon in Curbing Climate Change. *World Resources Institute*. https://www.wri.org/blog/2016/11/indigenous-peoples-and-local-communities-are-worlds-secret-weapon-curbing-climate#:~:text=Curbing%20Climate%20Change-,Indigenous%20Peoples%20and%20Local%20Communities%20Are%20the,Weapon%20in%20Curbing%20Climate%20Change&text=Some%20estimates%20suggest%20that%20as,of%20Indigenous%20and%20Community%20Lands.)

Rights and Resources Institute, et al. (2018). *A Global Baseline of Carbon Storage in Collective Lands: Indigenous and Local Community Contributions to Climate Change Mitigation*. Rights and Resources Institute. https://rightsandresources.org/wp-content/uploads/2018/09/A-Global-Baseline_RRI_Sept-2018.pdf

Risbey, J. S., & O'Kane, T. J. (2011). Sources of knowledge and ignorance in climate research. *Climatic Change*, *108*(4), 755. https://doi.org/10.1007/s10584-011-0186-6

Roychoudhury, A., Shepardson, D.P., & Hirsch, A.S. (2017). System Thinking and Teaching in the Context of Climate System and Climate Change. In *Teaching and Learning About Climate Change: A Framework for Educators* (First). Routledge.

Running, S. W. (2007). *The 5 Stages of Climate Grief*. 3.

Schwartz, J. (2020, January 17). Court Quashes Youth Climate Change Case Against Government. *New York Times*.

Shapin, Steven. (2018). *The Scientific Revolution* (Second). University of Chicago Press.

Author. (2021). Manuscript under preparation.

Author. (2015).

Solomon, F.C., Wright, T., Steele, M., & Champion, D. (2019). *Embodied Learning through Dance and Physics | TERC*. https://www.terc.edu/embodied-learning-through-dance-and-physics/

Steffen, W., Richardson, K., Rockström, J., Cornell, S. E., Fetzer, I., Bennett, E. M., Biggs, R., Carpenter, S. R., de Vries, W., de Wit, C. A., Folke, C., Gerten, D., Heinke, J., Mace, G. M., Persson, L. M., Ramanathan, V., Reyers, B., & Sörlin, S. (2015). Planetary boundaries: Guiding human development on a changing planet. *Science*, *347*(6223). https://doi.org/10.1126/science.1259855

Steffen, W., Rockström, J., Richardson, K., Lenton, T. M., Folke, C., Liverman, D., Summerhayes, C. P., Barnosky, A. D., Cornell, S. E., Crucifix, M., Donges, J. F., Fetzer, I., Lade, S. J., Scheffer, M., Winkelmann, R., & Schellnhuber, H. J. (2018). Trajectories of the Earth System in the Anthropocene. *Proceedings of the National Academy of Sciences*, *115*(33), 8252–8259. https://doi.org/10.1073/pnas.1810141115

Sterling, S. (2011). Transformative Learning and Sustainability: Sketching the conceptual ground. *Learning and Teaching in Higher Education*, *5*, 17–33.

Stuecker, M. F., Bitz, C. M., Armour, K. C., Proistosescu, C., Kang, S. M., Xie, S.-P., Kim, D., McGregor, S., Zhang, W., Zhao, S., Cai, W., Dong, Y., & Jin, F.-F. (2018). Polar amplification dominated by local forcing and feedbacks. *Nature Climate Change*, *8*(12), 1076–1081. https://doi.org/10.1038/s41558-018-0339-y

Sutter, John D. (2017, August 11). What Killed Stacy Ruffin? *Cnn.Com*. https://edition.cnn.com/2017/08/11/health/sutter-louisiana-flood-stacy-ruffin/index.html

Tigue, K. (2019, October 11). "We Must Grow This Movement": Youth Climate Activists Ramp up the Pressure. *Inside Climate News*. https://insideclimatenews.org/news/10112019/climate-change-school-strike-protests-extinction-rebellion-future-greta-thunberg

Tollefson, J. (2020). How Hot Will Earth Get by 2100? *Nature*, *580*, 443–445. https://doi.org/10.1038/d41586-020-01125-x

UN News. (2019, June). *World Faces "Climate Apartheid" Risk, 120 more million in poverty: UN Expert*. Global Perspective, Human Stories. https://news.un.org/en/story/2019/06/1041261

UNEP. (2019). *Global Resources Outlook 2019: Natural Resources for the Future We Want*. United Nations Environmental Program. https://wedocs.unep.org/bitstream/handle/20.500.11822/27518/GRO_2019_SPM_EN.pdf?sequence=1&isAllowed=y

US Global Change Research Program (Archived). (2009). *800,000 Year record of CO2 Concentration | Global Climate Change Impacts in the United States*. https://nca2009.globalchange.gov/global-climate-change/index.html

Wade Institute for Science Education. (2017, September). *Educators Explore Climate Change Concepts During MITS MetroWest Summer Institute*. https://www.wadeinstitutema.org/educators-explore-climate-change-concepts-mits-metrowest-summer-institute/

WEF Global Risks Report. (2017). The matrix of top 5 risks from 2007 to 2017. *Global Risks Report 2017*. http://wef.ch/2izWcoa

Weissberg, R. P., Durlak, J. A., Domitrovich, C. E., & Gullotta, T. P. (Eds.). (2015). Social and emotional learning: Past, present, and future. In *Handbook of social and emotional learning: Research and practice.* (pp. 3–19). The Guilford Press.

Weston, B. H. (2008). Climate Change and Intergenerational Justice: Foundational Reflections. *Vermont Journal of Environmental Law*, *9*(3), 375–430. JSTOR. https://doi.org/10.2307/vermjenvilaw.9.3.375

Wiedmann, T. O., Schandl, H., Lenzen, M., Moran, D., Suh, S., West, J., & Kanemoto, K. (2015). The material footprint of nations. *Proceedings of the National Academy of Sciences*, *112*(20), 6271–6276. https://doi.org/10.1073/pnas.1220362110

Wiek, A., Withycombe, L., & Redman, C. (2011). Key competencies in sustainability: A reference framework for academic program development. *Sustainability Science*, *6*, 203–218. https://doi.org/10.1007/s11625-011-0132-6

Yale Poorvu Center for Teaching and Learning. (n.d.). *Case-Based Learning*. Retrieved July 24, 2020, from https://poorvucenter.yale.edu/faculty-resources/strategies-teaching/case-based-learning

Yeager, D., Walton, G., & Cohen, G. L. (2013). Addressing Achievement Gaps with Psychological Interventions. *Phi Delta Kappan*, *94*(5), 62–65. https://doi.org/10.1177/003172171309400514

| Physics topic | Climate change topic |
|---|---|
| Contextualizing discussion (see next section for details) | Introduce Figure 1; elicit basic identification of Earth subsystems |
| thermal energy, temperature | Blackbody radiation, Stefan-Boltzmann law, radiative equilibrium, setting up equation for equilibrium surface temperature of a planet; calculation for Mars, Venus and Earth assuming no atmosphere; thermal expansion of ocean water leading to sea level rise |
| Fluids; flotation, laminar flow | Causes of sea level rise; ocean currents and their role in climate |
| Oscillations and resonance | Why carbon dioxide 'traps' infra-red radiation – resonance and the bending mode of vibration of $CO_2$; absorption spectrum of $CO_2$ |
| EM Spectrum | Connecting with earlier discussion on Blackbody radiation: the Sun's radiation spectrum; Natural greenhouse effect; Earth's radiation spectrum; Contrast planetary equilibrium temperatures from earlier discussion with actual surface temperatures for the inner planets; discrepancy is the greenhouse effect |
| Climate Week | The Big Picture: connecting climate concepts via three meta-concepts: Balance/Imbalance, Limits, Complexity (see below for details)<br>Introduction to complex systems and connection to climate change<br>Evidence, Impacts and climate justice<br>What needs to be done?<br>What are the barriers?<br>Start interdisciplinary project (optional) that integrates science and other aspects of climate in a real world context |
| Electromagnetism | Oscillating dipoles and understanding more fully why $CO_2$ is a greenhouse gas (optional) |
| End of course | Optional project completed |

TABLE 1: Integrating climate change topics into a general physics course.

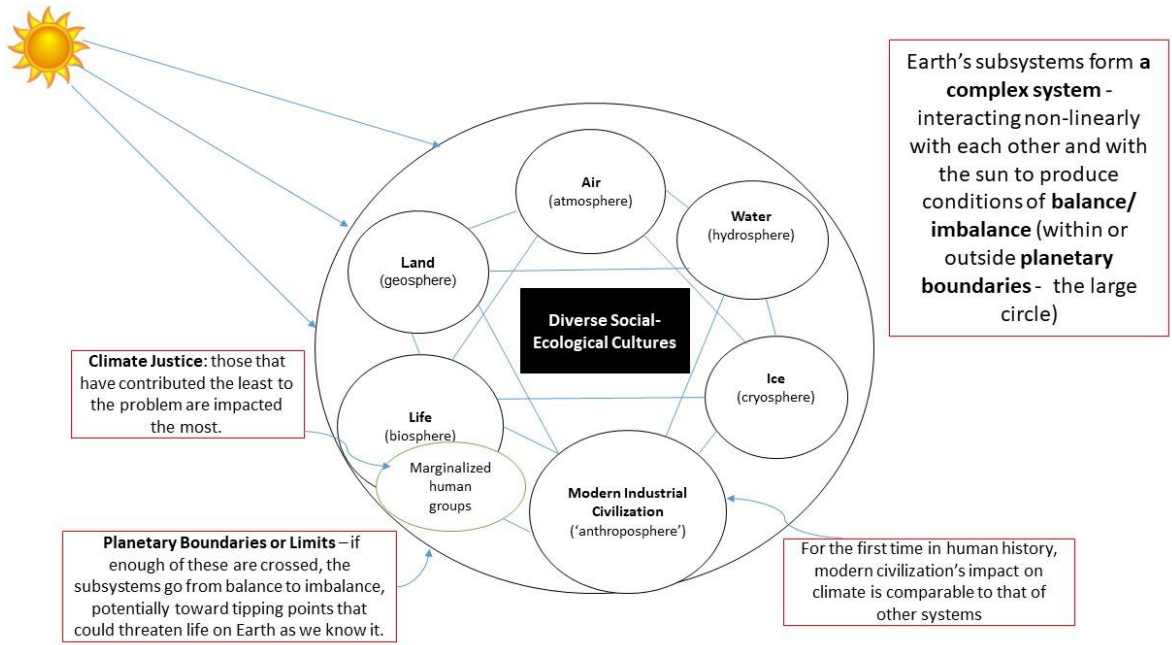

Figure 1: The three meta-concepts of Balance, Limits, and Complexity, along with Climate Justice

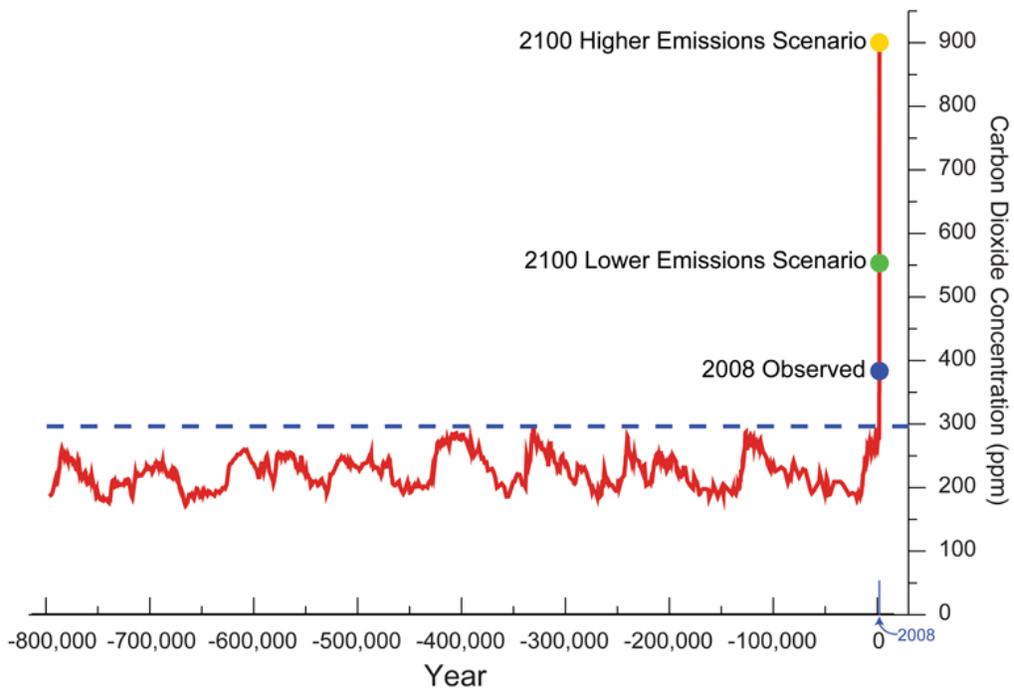

Figure 2: Historical Carbon Dioxide Concentration in the Atmosphere. Source: US Global Change Program (Archived) 2009.

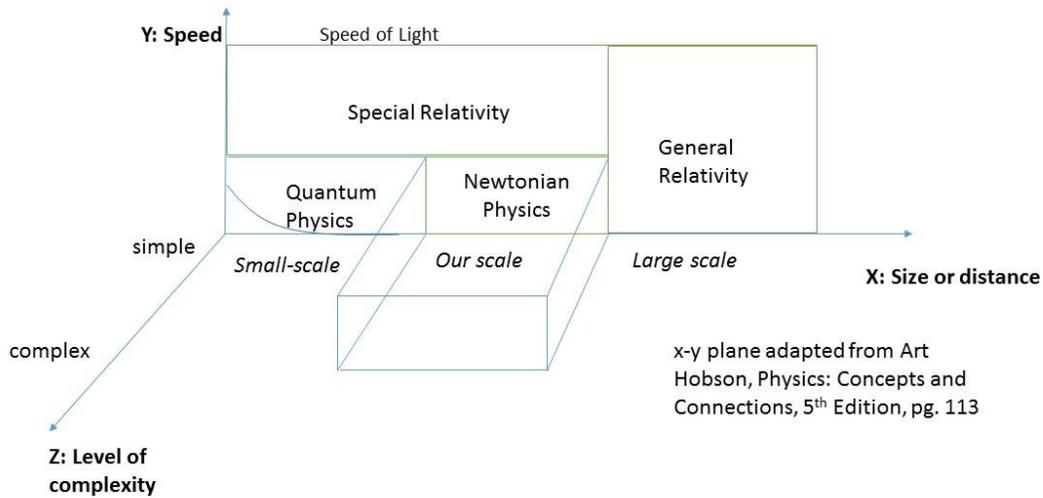

Figure 3: The domain of validity of Newtonian Physics

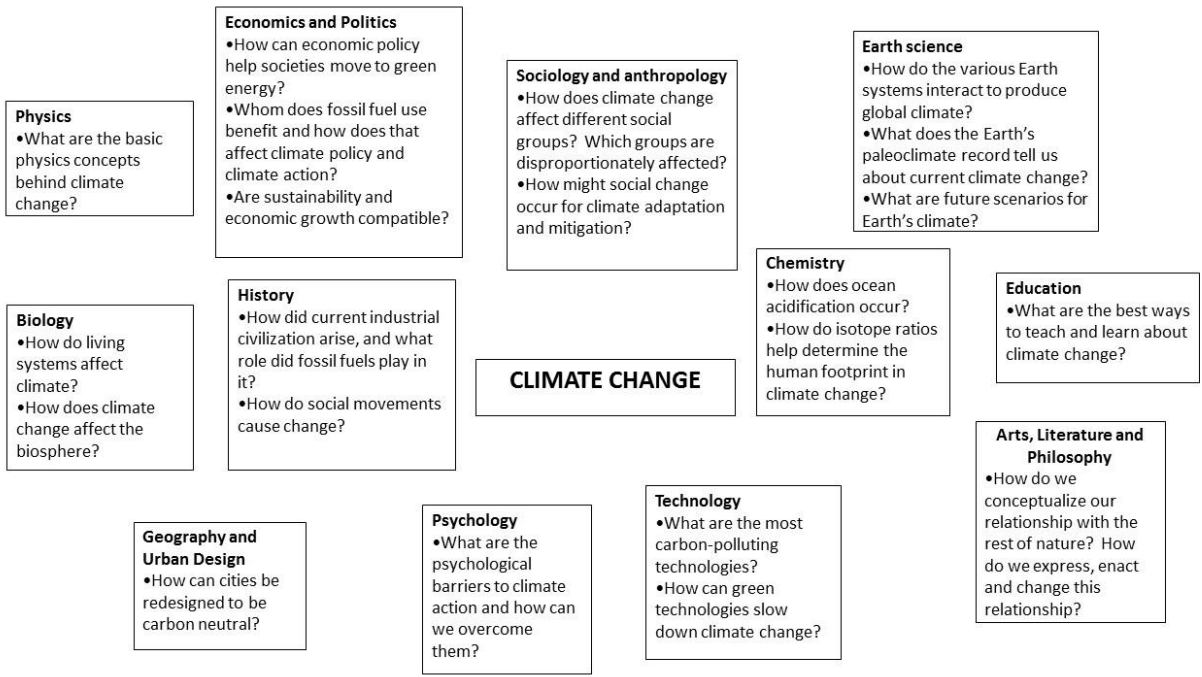

Figure 4: Climate Change involves multiple disciplines. A sampling of questions from different disciplinary perspectives.

**Figure Captions:**

Figure 1**:** The three meta-concepts of Balance, Limits, and Complexity, along with Climate Justice

Figure 2: Historical Carbon Dioxide Concentration in the Atmosphere. Source: US Global Change Program (Archived) 2009.

Figure 3: The domain of validity of Newtonian Physics

Figure 4: Climate Change involves multiple disciplines. A sampling of questions from different disciplinary perspectives.